\documentclass[ reprint,nofootinbib,amsmath,amssymb,
onecolumn,
aps]{revtex4-2}
\usepackage{appendix}
\usepackage{graphicx}
\usepackage{dcolumn}
\usepackage[colorlinks,linkcolor=magenta,anchorcolor=cyan,citecolor=blue]{hyperref}
\usepackage{physics}
\usepackage{bm}

\def\be{\begin{equation}}
	\def\ee{\end{equation}}
\def\ba{\begin{eqnarray}}
	\def\ea{\end{eqnarray}}

              \def\.{\cdot}

\begin{document}
	\title{Tidal Love Numbers of Thin-Shell Wormholes, de Sitter Gravastars, and Anti-de Sitter Gravastars via Full Perturbed Junction Conditions}
	\author{Gerui Chen$^{1,2,6}$}
	\email{t20152277@csuft.edu.cn}
	\author{Yu Tian$^{3,4}$}
	\email{ytian@ucas.ac.cn}
	\author{Hongbao Zhang$^{2,5}$}
	\email{hongbaozhang@bnu.edu.cn}

	\affiliation{$^1$ College of Electronic Information and Physics, Central South University of Forestry and Technology, Changsha 410004, China\\
	$^2$School of Physics and Astronomy, Beijing Normal University, Beijing 100875, China\\
	$^3$School of Physics, Chinese Academy of Sciences, Beijing 100190, China\\
	$^4$Institute of Theoretical Physics, Chinese Academy of Sciences, Beijing 100190, China\\
	$^5$Key Laboratory of Multiscale Spin Physics, Ministry of Education, Beijing Normal University, Beijing 100875, China\\
	$^6$Hunan Province Key Laboratory of Materials Surface/Interface Science and Technology, Central South University of Forestry and Technology, Changsha 410004, China}

	\begin{abstract}

		In this work, we revisit the tidal Love numbers (TLNs) of several classes of exotic compact objects (ECOs) using the recently derived full perturbed junction conditions. We investigate four representative thin-shell models: wormholes, de Sitter gravastars with zero surface energy density, general de Sitter gravastars, and anti-de Sitter gravastars. Previous studies of wormholes and de Sitter gravastars with zero surface energy density, based on simplified matching conditions, suggested that the TLNs of ultracompact ECOs vanish logarithmically in the black-hole (BH) limit, leading to the conjecture that this logarithmic dependence is a 
		generic feature of ultracompact exotic objects. We show that this logarithmic behavior is not universal once the full perturbed junction conditions are imposed. In particular, the odd-parity TLNs exhibit polynomial scaling in the BH limit for all the models considered, while the even-parity sector displays both logarithmic and polynomial behaviors depending on the model. We also find that the signs of the TLNs can differ qualitatively from previous results, especially in the odd-parity sector. Our results demonstrate that the tidal response of ECOs is strongly affected by the perturbed matching conditions and that the previously conjectured universal logarithmic scaling near the BH limit is not generic.

	\end{abstract}
	\maketitle
	\section{Introduction}
	
Einstein's general theory of relativity, which describes gravity as a manifestation of curved spacetime, predicts the existence of black holes~\cite{Wald,Chandrasekhar}. The existence of black holes as astrophysical objects has been firmly established through observations of gravitational waves, horizon-scale imaging, and the dynamics of galactic nuclei~\cite{B1,B2,R1,R2,AG,S,K,RS}. As observations probe increasingly subtle aspects of black-hole dynamics and
	theoretical predictions reach comparable precision, black holes serve as a unique laboratory for testing general relativity in the strong-field regime and exploring the nature of spacetime. However, black holes, as classical solutions of Einstein gravity, give rise to several puzzles, such as the black hole information paradox, the connection between black-hole entropy and horizon area, and the existence of the singularity inside black holes, all of which reveal a profound conflict between quantum mechanics and general relativity~\cite{Hawking1}. Considerable effort has been devoted to interpreting and reconciling these puzzling aspects of black holes. Given the profound difficulties in formulating a consistent quantum version of black holes, it is valuable to investigate alternative scenarios that do not involve traditional black holes. 
	  
	  Numerous alternative compact-object models have been proposed that can mimic many observational features of black holes while avoiding the presence of an event horizon or spacetime singularities. These objects, commonly referred to as black-hole mimickers, remain consistent with current observational constraints~\cite{V. Cardoso, M. Visser, N. K}. Any future observation—for example, involving ringdown quasinormal modes~\cite{Berti,Yang}—that favors a black-hole mimicker over a black hole would constitute a major breakthrough in fundamental physics. Horizonless compact objects constitute a prominent class of black-hole mimickers, which are collectively referred to as exotic compact objects (ECOs). Among them, objects such as boson stars~\cite{Jetzer1,Jetzer2,Jetzer3}, Dirac stars~\cite{Finster}, Proca stars~\cite{Brito}, and  fermion soliton stars~\cite{Lee,Grosso,Grosso1} possess smooth matter and field distributions, making them  suitable for stability studies and numerical investigations. However, constructing stable configurations that attain compactness comparable to that of black holes remains a nontrivial challenge. Another class of compact objects comprises configurations containing one or more thin membranes that separate distinct spacetime regions, such as thin-shell wormholes~\cite{ms} and gravastars~\cite{POM,mdl}. These constructions are particularly appealing because they allow the transition to an exterior spacetime identical to that of a black hole to occur arbitrarily close to the would-be horizon. As a result, such models can achieve compactness arbitrarily close to that of black holes. Moreover, compact objects with membranes are expected to exhibit strong-gravity dynamical behavior distinct from that of more uniform compact objects, which has motivated extensive studies of two-dimensional membranes.

	The forthcoming next-generation ground-based gravitational-wave (GW)
	detectors~\cite{Kalogera}, such as the Einstein Telescope~\cite{Maggiore,Branchesi} and
	Cosmic Explorer~\cite{Abbott,Essick}, will improve the accuracy of
	measurements of the tidal deformability~\cite{Pacilio,Forteza}. This improved accuracy could unveil new physics in the GW signals, potentially leading to the confirmation of the existence of these exotic compact objects. The deformability of a self-gravitating object immersed
	in an external tidal field is measured in terms of its TLNs~\cite{Love,Murray,Will}, which provide a robust analytical framework for understanding tidal effects. The TLNs encode the deformability of a self-gravitating object immersed in a tidal 
	environment and depend significantly both on the object’s internal structure and on the dynamics of the 
	gravitational field. Initially devised in the context of Newtonian gravity, the concept of TLNs has since been successfully extended to the full theory of general relativity~\cite{Hinderer,Poisson,Damour}, largely motivated by the prospect of measuring the TLNs of neutron stars (NSs)
	through GW detections and, in turn, understanding the
	behavior of matter at supranuclear densities~\cite{Lattimer1,Lattimer2,Lattimer3,Lattimer4,Lattimer5,Lattimer6,Lattimer7}. An intriguing result in classical  general relativity is the fact that the TLNs of a black hole are precisely zero. Tidal effects are also used to explore more fundamental questions related to
	the nature of compact objects and the behavior of gravity
	in the strong-field regime. In Ref.~\cite{Vitor}, the TLNs of several exotic compact objects, including wormholes, perfectly reflecting mirrors, and gravastars, were computed within general relativity.
	The authors found that the TLNs of these ECOs are generically nonvanishing. In the limit in which the compactness approaches the black-hole value, all TLNs vanish only logarithmically. Motivated by this result, they conjectured that such logarithmic scaling is a generic feature of ultracompact exotic objects and should hold for any ECO whose exterior geometry approaches that of a black hole in the $r_0\to 2M$ limit. Owing to this mild logarithmic behavior, the TLNs are not exceedingly small. Indeed, even in the Planckian case, $(r_0-2M)\sim l_p$, the TLNs have comparable magnitudes across different models and are only about five orders of magnitude smaller than those of a typical neutron star. Furthermore, all TLNs of these ultracompact exotic objects have the opposite sign relative to the neutron star case. 
	
    The TLNs of these models can be 
	computed in closed analytical form. The interior 
	and the exterior solutions are overall described by three 
	independent constants. Since the problem is linear, an 
	overall amplitude is irrelevant,
	and therefore only two junction conditions at $r_0$ were imposed in Ref.~\cite{Vitor}. Recently, the complete set of perturbed junction conditions was derived in Ref.~\cite{Cardoso1}.
	In view of the importance of tidal Love numbers, in this work, we revisit the tidal Love numbers of wormholes, de Sitter (dS) gravastars and anti-de Sitter (AdS) gravastars using these full perturbed junction conditions, and compare the resulting behaviors of TLNs with that obtained from the simplified junction conditions adopted in Ref.~\cite{Vitor}.
	
	The plan of the paper is as follows. In section 2, we discuss the geometry of the shell and the definition of tidal Love numbers in full General Relativity. In section 3,  we compute the tidal love numbers of wormholes, gravastars, and AdS bubbles. A brief summary is given in section 6.
	
	We use the geometric units $ G =  c = 1$.

\section{Tidal love numbers in General Relativity}
In this section, we give the definition of the tidal Love numbers in general relativity. We first discuss the geometry of the static, spherically symmetric shell. We then consider a perturbation to the shell, and give the definition of TLNs.

The static, spherically symmetric background metric is expressed in the following unified form,
\begin{eqnarray}
	ds_0^2=-f(r)dt^2+g(r)dr^2+r^2d\theta^2+r^2\sin^2\theta d\varphi^2.\label{q1}
\end{eqnarray}
To match the two spacetimes physically at the location of the shell, two junction conditions need to be imposed~\cite{Israel,Israel2,Eric}. The shell is described by parametric equations of the form
\begin{eqnarray}
	x_{\pm}^{\mu}=x_{\pm}^{\mu}(y^a),
\end{eqnarray}
where $y^a$ are the intrinsic coordinates of the shell, denoted as $y^a=(T,\Theta,\Phi)$, and $x_{\pm}^{\mu}$ refer to the coordinates used outside and inside the shell, respectively. The induced metric $\gamma_{ab}$ on the shell is defined by $\gamma_{ab}=g_{\mu\nu}e^{\mu}_{a}e^{\nu}_{b}$, where $e^\mu_a=\frac{\partial x^{\mu}}{\partial y^a}$ is a set of three independent tangent vectors to the shell. The unit normal vector to the shell, $n^{\mu}$, is another important quantity characterizing the shell, satisfying the following orthonormal relation:
\begin{eqnarray}
	n_{\mu}e^{\mu}_{a}=0, \;\; n^{\mu}n_{\mu}=1. \label{ne}
\end{eqnarray}

The first junction condition ensures that the induced metric $\gamma_{ab}$ is continuous across the shell, that is, $[[\gamma_{ab}]]=0$, where $[[...]]$ denotes the jump of a quantity across the shell. The second junction condition determines the stress-energy tensor on the thin shell in terms of  the jump in extrinsic curvature across it,
\begin{eqnarray}
	S_{ab}=-\frac{1}{8\pi}([[K_{ab}]]-\gamma_{ab}[[K]]),\label{j1}
\end{eqnarray}
where $K_{ab}=e^{\mu}_{a}e^{\nu}_{b}\nabla_{\mu}n_{\nu}$ is the extrinsic curvature, and $K=\gamma_{ab}K^{ab}$. This gives
\begin{eqnarray}
	S_{TT}&=&-\frac{1}{4\pi R}[[\frac{f}{\sqrt{g}}]],\label{923} \\
	S_{\Theta\Theta}&=&\frac{R}{8\pi}[[\frac{1}{\sqrt{g}}]]+\frac{R^2}{16 \pi}[[\frac{f'}{f\sqrt{g}}]],\label{922}
\end{eqnarray}
where the prime denotes a first-order radial derivative.

Let us now cast $S_{ab}$ in the form of a perfect fluid. The  stress-energy tensor of a perfect fluid is expressed as
\begin{eqnarray}
	S_{ab}=(\sigma+p)u_{a}u_{b}+p\gamma_{ab},\label{92}
\end{eqnarray}
in terms of a surface density $\sigma$, a surface pressure $p$, and a four-velocity field $u^a$ with only non-zero component $u^T=\frac{1}{\sqrt{f(R)}}$.
Combining Eqs.~(\ref{923}), (\ref{922}) and (\ref{92}), the surface-energy density $\sigma$ and pressure $p$ are
\begin{eqnarray}
	\sigma&=&-\frac{1}{4\pi R}[[\frac{1}{\sqrt{g}}]],\nonumber\\
	p&=&\frac{1}{8\pi R}[[\frac{1}{\sqrt{g}}]]+\frac{1}{16\pi}[[\frac{f'}{f\sqrt{g}}]].\label{dp}
\end{eqnarray}

In order to compute tidal Love numbers, it is necessary to derive expressions that relate the induced mass and current multipole moments to the external tidal field. This is achieved by applying linear perturbation theory to introduce a small deformation to the spacetime metric,
\begin{eqnarray}
	g_{\mu\nu}=g^{(0)}_{\mu\nu}+h_{\mu\nu},\label{full}
\end{eqnarray}
where $g^{(0)}_{\mu\nu}$ is the metric given in Eq.~(\ref{q1}), and $h_{\mu\nu}$ is a small perturbation.

Owing to spherical symmetry, the perturbation $h_{\mu\nu}$ admits a decomposition into spherical harmonics and a separation into even- and odd-parity sectors, $h_{\mu\nu} = h_{\mu\nu} ^{\text{even}}+h_{\mu\nu} ^{\text{odd}}$.
In the Regge-Wheeler gauge~\cite{Wheeler}, the metric perturbations are decomposed as
\begin{equation}
	h_{\mu\nu} ^{\text{even}}=
	\begin{pmatrix}
		f(r)H_{0}^{lm}(t,r)Y^{lm} & H_{1}^{lm}(t,r)Y^{lm} & 0 & 0\\
		H_{1}^{lm}(t,r)Y^{lm} & g(r)H_{2}^{lm}(t,r)Y^{lm} & 0 & 0\\
		0 & 0 & r^2 K^{lm}(t,r)Y^{lm} & 0\\
		0 & 0 & 0 & r^2 \sin^2\theta K^{lm}(t,r)Y^{lm}\\
	\end{pmatrix},
\end{equation}
\begin{equation}
	h_{\mu\nu} ^{\text{odd}}=
	\begin{pmatrix}
		0 & 0 & h_{0}^{lm}(t,r)S_{\theta}^{lm} & h_{0}^{lm}(t,r)S_{\varphi}^{lm}\\
		0 & 0 & h_{1}^{lm}(t,r)S_{\theta}^{lm} & h_{1}^{lm}(t,r)S_{\varphi}^{lm}\\
		h_{0}^{lm}(t,r)S_{\theta}^{lm} & h_{1}^{lm}(t,r)S_{\theta}^{lm} & 0 & 0\\
		h_{0}^{lm}(t,r)S_{\varphi}^{lm} & h_{1}^{lm}(t,r)S_{\varphi}^{lm} & 0 & 0\\
	\end{pmatrix},
\end{equation}
where, for brevity, $Y^{lm}\equiv Y^{lm}(\theta,\varphi)$, $S_{\theta}^{lm}\equiv -Y_{, \varphi}^{lm}/\sin\theta$, and $S_{\varphi}^{lm}\equiv\sin\theta Y_{, \theta}^{lm}$. There is an implicit summation over the angular indices ($l,m$). The perturbations $H_{0}^{lm}$, $H_{1}^{lm}$, $H_{2}^{lm}$, and $K^{lm}$ belong to the even-parity sector, while $h_{0}^{lm}$ and $h_{1}^{lm}$ belong to the odd-parity sector. We consider perturbations induced by an external static tidal field. In the static background spacetime, the perturbations are consequently described by time-independent functions. Moreover, as described in Regge and Wheeler's paper~\cite{Wheeler}, there is no need to work with an arbitrary $m$, so we prefer to set $m=0$ with the advantage that the dependence on $\varphi$ completely disappears from the calculations. We begin with the following perturbations:
\begin{equation}
	h_{\mu\nu} ^{\text{even}}=
	\begin{pmatrix}
		f(r)H_{0}(r)Y(\theta) & H_{1}(r)Y(\theta) & 0 & 0\\
		H_{1}(r)Y(\theta) & g(r)H_{2}(r)Y(\theta) & 0 & 0\\
		0 & 0 & r^2 K(r)Y(\theta) & 0\\
		0 & 0 & 0 & r^2 \sin^2\theta K(r)Y(\theta)\\
	\end{pmatrix},\label{951}
\end{equation}
\begin{equation}
	h_{\mu\nu} ^{\text{odd}}=
	\begin{pmatrix}
		0 & 0 & 0 & h_{0}(r)\sin\theta Y'(\theta)\\
		0 & 0 & 0 & h_{1}(r)\sin\theta Y'(\theta)\\
		0& 0 & 0 & 0\\
		h_{0}(r)\sin\theta Y'(\theta) & h_{1}(r)\sin\theta Y'(\theta) & 0 & 0\\
	\end{pmatrix},\label{952}
\end{equation}
where $Y(\theta) \equiv Y^{l0}(\theta)$, $Y'(\theta)\equiv\frac{dY(\theta)}{d\theta}$, and the superscript $l$ on the radial functions (e.g., $H_0$, $K$, etc.) has been omitted for brevity. Since the background spacetime is spherically symmetric, the two parity sectors decouple and can therefore be solved independently. By solving the linearized field equations for a specific model, one can obtain the metric perturbations given in Eqs.~(\ref{951}) and (\ref{952}). The explicit solutions are presented in Appendix~\ref{solution}.

Consider an isolated,
self-gravitating compact object immersed in a tidal environment. Following Ref.~\cite{Poisson},
the symmetric and trace-free even and odd tidal multipole moments of order $l$ are defined as $\mathcal{E}_{a_{1}...a_{l}}\equiv [(l-2)!]^{-1} \langle C_{0 a_1 0 a_2 ; a_3 \dots a_l} \rangle$ and $\mathcal{B}_{a_1 \dots a_l} \equiv \left[ \frac{2}{3} (l+1) (l-2)! \right]^{-1} \langle \epsilon_{a_1 b c} C^{b c}_{a_2 0; a_3 \dots a_l} \rangle$, where $C_{abcd}$ refers to the Weyl tensor, a semicolon denotes a covariant derivative, $\epsilon_{abc}$ is the permutation symbol, and the angular brackets denote symmetrization of the indices $a_{i}$ and the removal of all traces. The moments $\mathcal{E}_{a_{1}...a_{l}}$ (or $\mathcal{B}_{a_1 \dots a_l} $) can be decomposed into a basis of even- (or odd-) parity spherical harmonics. We denote by $\mathcal{E}^{lm}$ and $\mathcal{B}^{lm}$ the amplitudes of the even- and odd-parity components of the external tidal field with harmonic indices $(l,m)$, where $m$ is the azimuthal number ($|m|\leq l$).
The structure of the external tidal field is entirely encoded in the coefficients $\mathcal{E}^{lm}$ and $\mathcal{B}^{lm}$. Due to the external perturbations, 
the mass and current multipole moments ($M_l$ and $S_l$, respectively)
of the compact object are deformed. In linear perturbation
theory, these deformations are proportional to the applied
tidal field. The remaining task is to extract the multipole moments and tidal fields from the asymptotic behavior of the full spacetime
metric given in Eq.~(\ref{full}). Geroch and Hansen first provided a definition of multipole moments for axisymmetric and asymptotically flat spacetimes~\cite{Geroch,Hansen}. Then, in 1980, Thorne introduced an alternative approach to defining multipole coefficients of any spacetime metric expressed in asymptotically Cartesian and mass-centered coordinates~\cite{Thorne}. 
It has been demonstrated that these two formulations of multipole moments are equivalent~\cite{YG}. The multipole moments can be extracted from the asymptotic behavior of the spacetime metric,
\begin{eqnarray}
	g_{tt}&=&-1+\frac{2M}{r}+\sum_{l=2}^{\infty} \left(\frac{2}{r^{l+1}}\left[\sqrt{\frac{4\pi}{2l+1}}M_{l}Y^{l0}+(l'<l \, \text{pole})\right]
	-\frac{2}{l(l-1)}r^{l}\left[\mathcal{E}_l Y^{l0}+(l'<l \,\text{pole}\right ]\right),\label{gtt}
\end{eqnarray}
\begin{eqnarray}
	g_{t\varphi}&=&\frac{2J}{r}\sin^2\theta+\sum_{l=2}^{\infty} \left(\frac{2}{r^l}\left[\sqrt{\frac{4\pi}{2l+1}}\frac{S_l}{l}S^{l0}_{\phi}+
	(l'<l \,\text{pole})\right]+\frac{2r^{l+1}}{3l(l-1)}\left[\mathcal{B}_l S^{l0}_{\varphi}+(l'<l \,\text{pole}) \right]\right),\label{gtb}
\end{eqnarray}
where $\mathcal{E}_l$ and $\mathcal{B}_l$ are, respectively, the amplitudes of the even- and odd-parity components of the external field with harmonic number $l$. The condition $m=0$ is set by imposing axisymmetry. 
The even-parity and odd-parity tidal Love numbers are defined as~\cite{Vitor}
\begin{eqnarray}
	k_l^E&=&-\frac{1}{2}\frac{l(l-1)}{M^{2l+1}}\sqrt{\frac{4\pi}{2l+1}}\frac{M_l}{\mathcal{E}_l},\label{ke}\\
	k_l^B&=&-\frac{3}{2}\frac{l(l-1)}{(l+1)M^{2l+1}}\sqrt{\frac{4\pi}{2l+1}}\frac{S_l}{\mathcal{B}_{l}}\label{kb},
\end{eqnarray}
where $M$ is the mass of the compact object.

For odd-parity perturbations, as $r \rightarrow \infty$, the asymptotic behavior of the exterior solution (\ref{oddout}) is
\begin{eqnarray}
	h_0(r)&\sim& \tilde{c}_1(\frac{r}{M})^{l+1}+\tilde{c}_2(\frac{M}{r})^l\nonumber\\
	&=&(\frac{r}{M})^{l+1}\tilde{c}_1\left( 1+M^{2l+1}\frac{\tilde{c}_2}{\tilde{c}_1}\frac{1}{r^{2l+1}}\right).
\end{eqnarray}
By comparing this asymptotic behavior with Eq.~(\ref{gtb}) and using the definition of the odd-parity TLN in Eq.~(\ref{kb}), we obtain
\begin{eqnarray}
	k_{l}^B=-\frac{1}{2}\frac{l}{l+1}\frac{\tilde{c}_2}{\tilde{c}_1}. \label{eq1}
\end{eqnarray}
For even-parity perturbations, as $r \rightarrow \infty$, the asymptotic behavior of the exterior solution (\ref{evenout}) is
\begin{eqnarray}
	H_{\text{out}}\sim c_1(\frac{r}{M})^l+c_2(\frac{M}{r})^{l+1}\nonumber\\
	=c_{1}(\frac{r}{M})^l\left( 1+M^{2l+1}\frac{c_2}{c_1}\frac{1}{r^{2l+1}} \right).
\end{eqnarray}
Comparing this asymptotic behavior with Eq.~(\ref{gtt}) and using the definition of the even-parity TLN in Eq.~(\ref{ke}), we obtain
\begin{eqnarray}
	k_{l}^E=\frac{1}{2}\frac{c_2}{c_1}\label{eq2}.
\end{eqnarray}
We will use Eqs.~(\ref{eq1}) and (\ref{eq2}) to compute the tidal Love numbers in the next section.

\section{Tidal love numbers and their behavior in the black hole limit}

In this section, we compute the tidal Love numbers of wormholes, de Sitter gravastars, and anti-de Sitter gravastars by imposing the full perturbed junction conditions, and compare the results with those obtained in Ref.~\cite{Vitor}. Since some expressions for generic compactness are cumbersome, we present explicit formulas only in the BH limit.

\subsection{Wormholes}
The simplest wormhole models are constructed by taking two copies of the Schwarzschild solution and excising from them the four-dimensional regions defined by $r_{1,2}\leq r_0$~\cite{Visser}. This procedure yields two manifolds whose geodesics terminate on the timelike hypersurfaces
\begin{eqnarray}
\partial \Omega_{1,2} \equiv \{r_{1,2}=r_0  \mid r_0>2 M\}.
\end{eqnarray}	
The two copies are then glued together by identifying these two boundaries, i.e., $\partial\Omega_{1} =\partial\Omega_{2}$, so that the resulting spacetime is geodesically complete and consists of two distinct regions connected by a wormhole with a throat at $r=r_0$. The patching at the throat requires a thin shell of matter with the surface energy density and surface pressure,
\begin{eqnarray}
	\sigma=-\frac{1}{2\pi r_0}\sqrt{1-\frac{2 M}{r_0}}, \qquad	p=\frac{1}{4 \pi r_0}\frac{1-M/r_0}{\sqrt{1-2 M/r_0}}.
\end{eqnarray}

We first review the tidal Love numbers obtained in Ref.~\cite{Vitor}. By requiring regularity of the odd-parity perturbations
across the shell, Ref.~\cite{Vitor} assumes that $h_0$ and its derivative with
respect to the radial tortoise coordinate $r_*$ are smooth. Thus, in the odd-parity sector, the matching conditions for wormholes and gravastars are
\begin{eqnarray}
[[h_0]]=0, \; \;  \; [[dh_0/dr_{*}]]=0.\label{jc1}
\end{eqnarray}
By imposing these two junction conditions, 
the odd-parity TLNs for arbitrary $l$ can be derived. The full analytic expressions for the odd-parity TLNs with $l=2$ and $l=3$ are presented in Ref.~\cite{Vitor}. In the BH limit,
the odd-parity TLNs of wormholes behave as~\cite{Vitor}
\begin{eqnarray}
	k_2^{B}&\sim&\frac{16}{5(31+12 \log \xi)},\nonumber\\
k_3^{B}&\sim&\frac{16}{7(209+ 60 \log \xi)},
\end{eqnarray}
where $\xi \equiv r_0/(2M)-1$, and the subleading terms of $\mathcal{O}(\frac{\xi}{(\log{\xi})^2})$ have been omitted.

For polar perturbations, Ref.~\cite{Vitor} uses the formalism developed in Ref.~\cite{Paolo1} to derive the junction conditions $[[K]]=0$ and $[[dK/dr_{*}]]=-8 \pi \sqrt{f(r_0)} \delta\Sigma$, where $\delta \Sigma$ is the perturbation of the surface energy density
of the thin shell. For simplicity, they assume that the thin-shell material is stiff, 
so that $\delta \Sigma \sim 0$. Therefore, in the polar sector for the wormhole and
gravastar cases, Ref.~\cite{Vitor} imposes the following conditions:
\begin{eqnarray}
	[[K]]=0, \; \;  [[dK/dr_{*}]]=0.\label{jc2}
\end{eqnarray}
 The even-parity TLNs for generic compactness are also  provided in Ref.~\cite{Vitor}.\footnote{There is a typo in the expression for $k_3^E$: the term $\log(\frac{1}{1+\xi}-1)$ should read $\log(1-\frac{1}{1+\xi})$.} In the BH limit, the even-parity TLNs of wormholes behave as~\cite{Vitor}
\begin{eqnarray}
    k_2^{E}&\sim&\frac{4}{5(8+ 3\log \xi)}, \nonumber\\
    k_3^{E}&\sim&\frac{8}{105(7 + 2\log \xi)}\label{kw1}.
\end{eqnarray}
The logarithmic dependence of the TLNs is particularly interesting because it implies that the deviations from zero (i.e., from the BH case) are relatively large even when the throat is located just a Planck length away from the would-be horizon.

Next, we revisit the tidal Love numbers of wormholes by imposing the full perturbed junction conditions presented in Appendix~\ref{jc}. These conditions are derived from the first and second junction conditions without any approximations.

For wormholes, the matching conditions for axial perturbations are
\begin{eqnarray}
0&=&[[h_0]],\\
8 \pi \sigma h_{0\text{out}} & = &\frac{1}{2} \frac{h^\prime_{0\text{out}}}{\sqrt{g_{\text{out}}}}+\frac{1}{2}\frac{h^\prime_{0\text{in}}}{\sqrt{g_{\text{in}}}}-\frac{2}{r_0}\frac{1}{\sqrt{g_{\text{out}}}}h_{0\text{out}}\nonumber\\
&-&\frac{2}{r_0}\frac{1}{\sqrt{g_{\text{in}}}}h_{0\text{in}}
		-\frac{1}{2}\frac{f^{\prime}_{\text{out}}}{f_{\text{out}}\sqrt{g_{\text{out}}}}h_{0\text{out}}-\frac{1}{2}\frac{f^{\prime}_{\text{in}}}{f_{\text{in}}\sqrt{g_{\text{in}}}}h_{0\text{in}},\label{no2}
\end{eqnarray}
where the first equation corresponds to Eq.~(\ref{1jc}), and the second equation follows from 
Eq.~(\ref{odd2}).\footnote{The extrinsic curvatures on the two sides of the wormhole have opposite signs. Therefore, the notation $[[f]]$ appearing in 
	the second junction condition should be understood as 
	$f_{\text{out}}+f_{\text{in}}$.} Equation (\ref{no2}) differs from the second equation in Eqs.~(\ref{jc1}). By imposing these two junction conditions, we obtain the odd-parity TLNs for  $l = 2$ and $l = 3$. 
	In the BH limit, they behave as
\begin{eqnarray}
	k_2^B &\sim&	\frac{16}{15}\xi-\frac{32}{45} (11 + 6 \log \xi)\xi ^2+\mathcal{O}\left(\xi
	^3\right),\nonumber\\
k_3^B &\sim&	\frac{8}{21}\xi -\frac{20}{63} (31 + 12 \log \xi )\xi^2 + \mathcal{O}\left(\xi ^3\right).
\end{eqnarray}
The odd-parity TLNs derived from the full perturbed junction conditions also vanish in the BH limit, $r_0\to2M$.

\begin{figure}[htbp]
	\centering
	\includegraphics[width=0.5\textwidth]{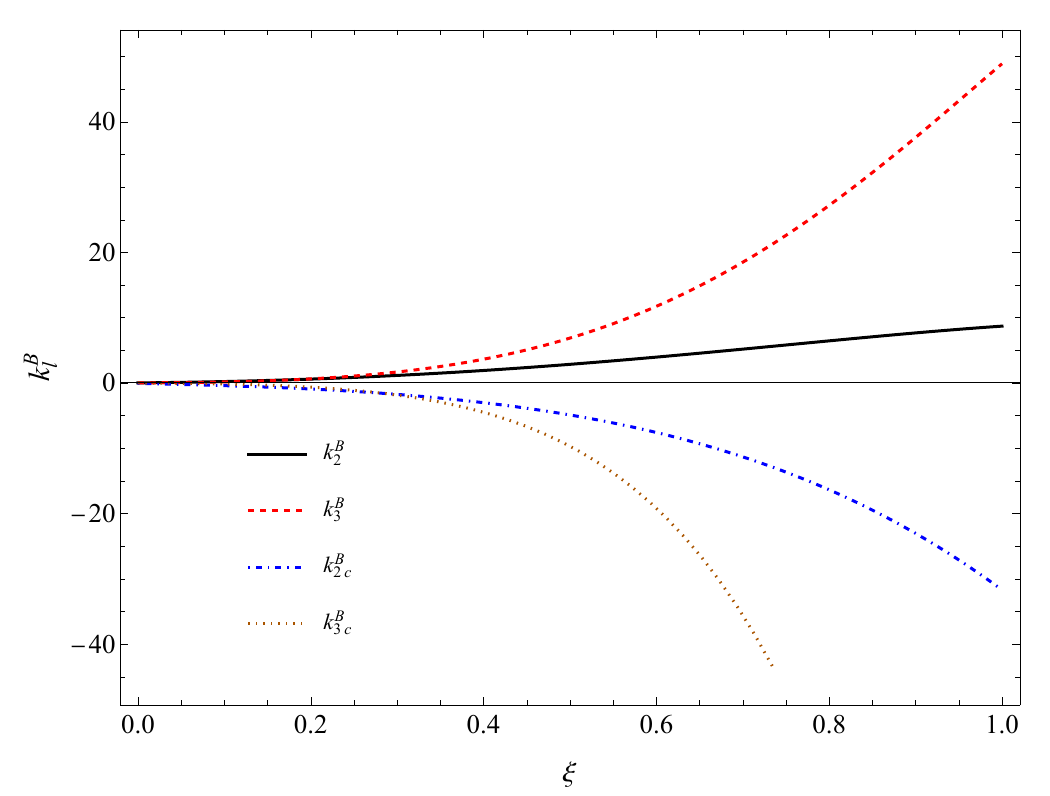}
	\caption{The odd-parity TLNs of thin-shell wormholes for $l=2$ and $l=3$. The odd-parity TLNs obtained from the two different sets of perturbed junction conditions are shown together as functions of the compactness $\xi$. $k_{2c}^B$ and $k_{3c}^B$ correspond to the odd-parity TLNs obtained from the simplified junction conditions, while $k_{2}^B$ and $k_{3}^B$ correspond to those obtained from the full junction conditions. The TLNs all vanish in the BH limit, $\xi \rightarrow 0$. Moreover, $k_{2c}^B$ and $k_{3c}^B$ are negative, whereas $k_{2}^B$ and $k_{3}^B$ are positive in the region $0<\xi<1$.}
	\label{fig1}
\end{figure}

Compared with the odd-parity TLNs obtained using the simplified junction conditions in Ref.~\cite{Vitor}, several notable differences arise. First, in the BH limit, the odd-parity TLNs derived from the full perturbed junction conditions do not exhibit logarithmic behavior; instead, they display a polynomial dependence, with the leading-order term scaling as $\mathcal{O}(\xi)$. Second, the TLNs remain positive throughout the region $0<\xi<1$, whereas the corresponding results in Ref.~\cite{Vitor} are negative. Figure~\ref{fig1} shows the odd-parity TLNs as functions of the compactness for the two 
different sets of perturbed junction conditions.

In the even-parity sector, there are five matching conditions. The first junction condition yields Eqs.~(\ref{fjc2}),
\begin{eqnarray}
	 \frac{2}{r_0}[[\delta r]]=-[[K]], \; \; [[H_0]]=[[\frac{\delta r f^{\prime}}{f}]].
\end{eqnarray}

Equation~(\ref{even1}), derived from the second junction condition, implies
\begin{eqnarray}
	\sqrt{g_{\text{out}}}\delta r_{\text{out}} + \sqrt{g_{\text{in}}}\delta r_{\text{in}} = 0.
\end{eqnarray}

Equations~(\ref{even2}) and (\ref{even3}), which also follow from the second junction condition, yield
\begin{eqnarray}
	&&\frac{2}{r_0}\frac{H_{0\text{out}}}{\sqrt{g_{\text{out}}}}+
	\frac{1}{r_0}\frac{H_{2\text{out}}}{\sqrt{g_{\text{out}}}}-\frac{K^{\prime}_{\text{out}}}{\sqrt{g_{\text{out}}}}+	\frac{2}{r_0}\frac{H_{0\text{in}}}{\sqrt{g_{\text{in}}}}+
	\frac{1}{r_0}\frac{H_{2\text{in}}}{\sqrt{g_{\text{in}}}}-\frac{K^{\prime}_{\text{in}}}{\sqrt{g_{\text{in}}}}\nonumber\\
	&&=8\pi \delta \sigma + 8\pi \sigma(\frac{f^{\prime}_{\text{out}}\delta r_{\text{out}}}{f_{\text{out}}} -H_{0\text{out}}),
\end{eqnarray}
\begin{eqnarray}\label{even6}
	&&-\frac{1}{2r_0}\frac{H_{2\text{out}}}{\sqrt{g_\text{out}}}+\frac{1}{r_0}\frac{K_\text{out}}{\sqrt{g_\text{out}}}-\frac{1}{4}\frac{H_{2\text{out}}f^{\prime}_\text{out}}{f_\text{out}\sqrt{g_\text{out}}}-\frac{1}{2r_0}\frac{H_{2\text{in}}}{\sqrt{g_\text{in}}}+\frac{1}{r_0}\frac{K_\text{in}}{\sqrt{g_\text{in}}}-\frac{1}{4}\frac{H_{2\text{in}}f^{\prime}_\text{in}}{f_\text{in}\sqrt{g_\text{in}}}\nonumber\\
	&&+\frac{1}{2}\frac{K_\text{out}f^{\prime}_\text{out}}{f_\text{out}\sqrt{g_\text{out}}}+\frac{1}{2}\frac{K^{\prime}_\text{out}}{\sqrt{g_\text{out}}}-\frac{1}{2}\frac{H_{0\text{out}}^{\prime}}{\sqrt{g_\text{out}}}+\frac{1}{2}\frac{K_\text{in}f^{\prime}_\text{in}}{f_\text{in}\sqrt{g_\text{in}}}+\frac{1}{2}\frac{K^{\prime}_\text{in}}{\sqrt{g_\text{in}}}-\frac{1}{2}\frac{H_{0\text{in}}^{\prime}}{\sqrt{g_\text{in}}}\nonumber\\
	&&=8\pi \delta p+8\pi p(K_{\text{out}}+2\frac{\delta r_{\text{out}}}{r_0}). 
\end{eqnarray}
The junction conditions for even-parity perturbations are more complicated because they couple to the matter degrees of freedom on the thin shell.
A complete description requires introducing an equation of state that relates $\delta p$ to $\delta \sigma$,
\begin{eqnarray}
	\delta p = v_s^2 \delta \sigma,
\end{eqnarray}
where $v_s$ is the sound speed of the fluid. Using these five junction conditions, one obtains the even-parity TLNs for arbitrary $l$. Since the complete even-parity results depend on $v_s$ and the expressions are excessively complicated, we do not present them here. In the BH limit, the even-parity TLNs for $l=2$ and $l=3$ behave as
\begin{eqnarray}
	k_2^{E}&\sim& \frac{4}{5 (7 + 3 \log \xi)}+\frac{76 \xi }{5 (7 + 3 \log \xi)^2}+\mathcal{O}\left(\xi^2\right),\nonumber\\
k_3^{E}&\sim& \frac{4}{35 (10 + 3 \log \xi)}+ \frac{212 \xi }{35 (10 + 3 \log \xi)^2}+ \mathcal{O}\left(\xi ^2\right).\label{kw2}
\end{eqnarray}
The sound speed $v_s$ enters at order $\mathcal{O}\left(\xi^2\right)$.
Comparing Eqs.~(\ref{kw1}) and (\ref{kw2}), one finds that the even-parity TLNs derived from the full matching conditions still exhibit logarithmic behavior in the BH limit, although there are discrepancies in the constant coefficients between the two expressions.  

In summary, the TLNs of thin-shell wormholes derived from the full perturbed junction conditions exhibit significant differences from those obtained using the simplified junction conditions in Ref.~\cite{Vitor}. In the odd-parity sector, although the TLNs still vanish in the BH limit $r_0\to2M$, their asymptotic behavior changes qualitatively: instead of displaying logarithmic scaling, they exhibit a polynomial dependence with leading-order behavior of $\mathcal{O}(\xi)$, while remaining positive throughout the range $0<\xi<1$. By contrast, the odd-parity TLNs obtained in Ref.~\cite{Vitor} are negative and exhibit logarithmic behavior in the BH limit. In the even-parity sector, the TLNs derived from the full perturbed junction conditions continue to exhibit logarithmic behavior in the BH limit, although the constant coefficients differ from those obtained using the simplified perturbed junction conditions. These results demonstrate that the full perturbed junction conditions can substantially modify both the quantitative and qualitative properties of the TLNs of thin-shell wormholes, particularly in the odd-parity sector.

\subsection{De Sitter gravastars}

Gravastars were first introduced in 2001 by Mazur and Mottola~\cite{po}. The interior of
these objects is described by a patch of de Sitter space,
which is smoothly connected to the Schwarzschild exterior
through an intermediate region filled with a perfect fluid. A particularly simple model is the so-called thin-shell
gravastar~\cite{vdl}, in which the thickness of the intermediate
region shrinks to zero.  Since 
their introduction, substantial effort has been devoted to understanding these objects and determining how they could 
be distinguished from black holes through gravitational-wave measurements \cite{nk, ve}. In this section, we revisit the TLNs of dS thin-shell gravastars based on the full perturbed junction conditions. We first compute the TLNs of dS gravastars composed of a fluid with zero energy density and then consider more general dS gravastars.

\subsubsection{De Sitter gravastars  with zero energy density}

For dS gravastars with zero energy density, the background metric takes the form
\begin{equation}
	\begin{cases}
		f(r) = g(r)^{-1}=  (1 - 2M/r), \qquad \quad   r > r_0, \\
		f(r) = g(r)^{-1} = 1-\frac{1}{3}\Lambda  r^2, \qquad \quad  \quad r < r_0,
	\end{cases}
\end{equation}
where $M$ is the mass of the gravastar, and $\Lambda$ is the positive cosmological constant. The condition $\Lambda = 6 M/r_0^3$ yields zero energy density. The surface pressure is 
\begin{eqnarray}
	p=\frac{3 M}{8 \pi r_0^2}\frac{1}{\sqrt{1-2 M/r_0}}.
\end{eqnarray}
In this model, the thin shell is described by a fluid with zero energy density and positive pressure.

By imposing the junction conditions (\ref{jc1}, \ref{jc2}), Ref.~\cite{Vitor} derived the full analytic expressions for the TLNs with $l=2$ and $l=3$.\footnote{There is a typo in the expression for $k_2^B$ given in Ref.~\cite{Vitor}: the entire expression has an extra minus sign.} In the BH limit, the TLNs behave as
\begin{eqnarray}
	k_2^B \sim	\frac{32}{5(43-12 \log 2+ 18 \log \xi)},\nonumber
\end{eqnarray}
\begin{eqnarray}
	k_3^B \sim	\frac{32}{7(307-60 \log 2+ 90 \log \xi)},\label{kkb}
\end{eqnarray} 
\begin{eqnarray}
	k_2^E \sim	\frac{16}{5(23-6 \log 2+ 9 \log \xi)},\nonumber
\end{eqnarray}
\begin{eqnarray}
	k_3^E \sim	\frac{16}{35(31-6 \log 2+ 9 \log \xi)}.\label{k2e2}
\end{eqnarray}
In the gravastar case, the TLNs are also negative and vanish only logarithmically in the BH limit. 

\begin{figure}[htbp]
	\centering
	\includegraphics[width=0.5\textwidth]{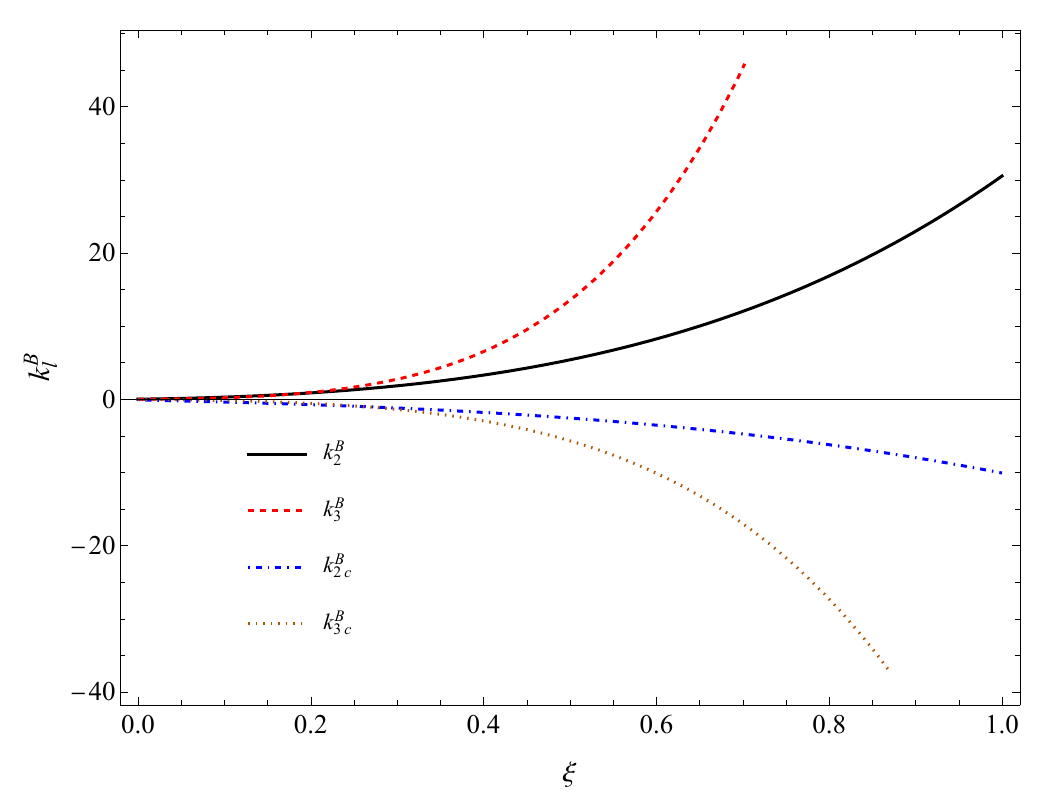}
	\caption{The odd-parity TLNs of dS gravastars for $l=2$ and $l=3$. The odd-parity TLNs obtained from the two different sets of perturbed junction conditions are shown together as functions of the compactness $\xi$. $k_{2c}^B$ and $k_{3c}^B$ correspond to the odd-parity TLNs obtained from the simplified junction conditions, while $k_{2}^B$ and $k_{3}^B$ correspond to those obtained from the full junction conditions. The TLNs all vanish in the BH limit. Moreover, $k_{2c}^B$ and $k_{3c}^B$ are negative, whereas $k_{2}^B$ and $k_{3}^B$ are positive in the region $0<\xi<1$.}
	\label{fig2}
\end{figure}

Instead, using the full perturbed junction conditions, we obtain the odd-parity TLNs for $l=2$ and $l=3$. 
In the BH limit, they behave as 
\begin{eqnarray}
	k_2^B \sim\frac{64 }{45}\xi+\frac{32}{405}  (-55+12 \log 2-54 \log \xi)\xi^2+ \mathcal{O}\left(\xi^3\right),\nonumber
\end{eqnarray}
\begin{eqnarray}
	k_3^B \sim \frac{32 }{63}\xi + \frac{16}{567} (-251 + 30 \log 2-135 \log \xi ) \xi^2+ \mathcal{O}\left(\xi^3\right).
\end{eqnarray}
The odd-parity TLNs obtained from the full perturbed junction conditions also vanish in the BH limit, $r_0 \rightarrow 2M$. However, they differ qualitatively from the TLNs in Eq.~(\ref{kkb}) derived from the simplified perturbed junction conditions. In particular, their asymptotic behavior near the BH limit is polynomial rather than logarithmic, with the leading contribution scaling as $\mathcal{O}(\xi)$. Moreover, the odd-parity TLNs remain positive throughout the entire interval $0<\xi<1$, in contrast to the previously reported negative values. The odd-parity TLNs of thin-shell gravastars with vanishing energy density, obtained from the two perturbed junction conditions, are shown as functions of the compactness in Fig.~\ref{fig2}.

For even parity, the behavior of the TLNs in the BH limit reads
\begin{eqnarray}
k_2^E \sim \frac{16}{5 (23 - 6 \log 2 + 9 \log \xi)}-\frac{24 (-76+30 \log 2-15 \log \xi)}{5 (-23+6
	\log 2-9 \log \xi)^2}\xi+\mathcal{O}\left(\xi ^2\right),\nonumber
\end{eqnarray}
\begin{eqnarray}
k_3^E \sim \frac{16}{35 ( 31 - 6 \log 2+9 \log \xi)}-\frac{72 (-68+22 \log 2-11 \log \xi )}{35 ( -31+6
	\log 2 -9 \log \xi)^2}\xi + \mathcal{O}\left(\xi ^2\right).\label{kge2}
\end{eqnarray}
The dependence on the sound speed $v_s$ appears at order $\mathcal{O}\left(\xi^2\right)$. The even-parity TLNs of dS gravastars with zero energy density, derived from the full matching conditions, also vanish and exhibit logarithmic behavior in the BH limit. Moreover, the leading-order terms in Eq.~(\ref{kge2}) coincide with those in Eq.~(\ref{k2e2}), while differences appear only at higher orders.
 
 \subsubsection{General de Sitter gravastars}

We now discuss the tidal Love numbers of general dS gravastars. The background metric is given by
\begin{equation}
\begin{cases}
	f(r) = g(r)^{-1}=  (1 - 2M/r), \qquad \quad \quad \quad \quad \quad \quad \quad \quad   r > r_0, \\
	f(r) = \frac{\left(1-2 M/R\right)}{1-\frac{\Lambda  }{3}R^2}(1-\frac{\Lambda}{3} r^2), \; \;
	
	g(r)^{-1} = 1-\frac{\Lambda}{3} r^2, \qquad  \,\, r < r_0. \label{gds}
\end{cases}
\end{equation}
Substituting the metric (\ref{gds}) into Eqs. (\ref{dp}), 
we obtain the surface energy density and surface pressure of de Sitter gravastars as
\begin{eqnarray}
	\sigma &=&\frac{1}{4\pi r_0}\left(\sqrt{1-\frac{r_0^2 }{3}\Lambda}-\sqrt{1-\frac{2 M}{r_0}}\right),\\
	 p &=& \frac{1}{24 \pi r_0}\left(\frac{2 r_0^2 \Lambda-3 }{\sqrt{1-r_0^2 \Lambda/3}}+3\frac{ 1-M/r_0}{\sqrt{1-2 M/r_0}}\right).
\end{eqnarray}

\begin{figure}[htbp]
	\centering
	\includegraphics[width=0.45\textwidth]{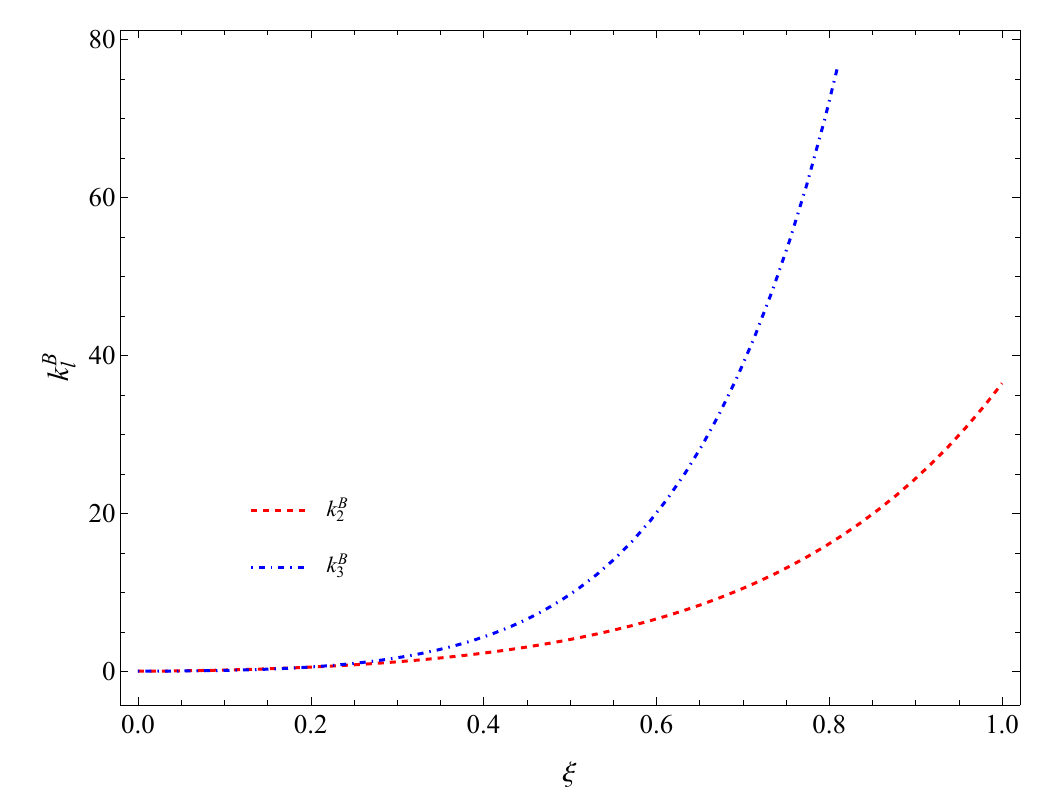}
	\caption{The odd-parity TLNs of dS gravastars with nonzero energy density for $l=2$ and $l=3$. The TLNs all vanish in the BH limit and are positive in the region $0<\xi<1$.}
	\label{fig3}
\end{figure}

To illustrate the numerical results, 
we follow the methods of Ref.~\cite{Uchikata1} and consider an equilibrium solution of a spherical gravastar characterized by the fixed value $M/L=1/5$. The de Sitter horizon radius $L=\sqrt{3/\Lambda}$ is set to unity, which implies $\Lambda=3$ and $M=1/5$. Using the full perturbed junction conditions, we compute the odd-parity TLNs for $l=2$ and $l=3$. In the black-hole limit, the odd-parity TLNs take the form
\begin{eqnarray}
		k_2^B  &\sim & \frac{32  \left[ 1775 \sqrt{3} \, _2F_1(\frac{1}{2},2;\frac{7}{2};\frac{4}{25})+48 \sqrt{3} \,
			_2F_1 (\frac{3}{2},3;\frac{9}{2};\frac{4}{25})\right]}{5625 \sqrt{7} \, _2F_1(\frac{1}{2},2;\frac{7}{2};\frac{4}{25})}\xi ^{3/2}+\mathcal{O}\left(\xi ^2\right)\nonumber\\	&\approx&	6.81271 \xi ^{3/2}+\mathcal{O}\left(\xi ^2\right),
		\end{eqnarray}
		\begin{eqnarray}
			k_3^B  &\sim &
			\frac{64 \left[345 \sqrt{3} \, _2F_1 (1,\frac{5}{2};\frac{9}{2};\frac{4}{25})+14 \sqrt{3} \,
				_2F_1 (2,\frac{7}{2};\frac{11}{2};\frac{4}{25})\right]}{4725 \sqrt{7} \, _2F_1(1,\frac{5}{2};\frac{9}{2};\frac{4}{25})}\xi^{3/2}+ \mathcal{O}\left(\xi ^2\right)\nonumber\\
				&\approx& 3.19969 \xi^{3/2}+\mathcal{O}\left(\xi ^2\right).
\end{eqnarray}
The odd-parity TLNs of general de Sitter gravastars also vanish as a power law in the BH limit, with the leading-order term scaling as $\mathcal{O}\left(\xi^{3/2}\right)$. The odd-parity TLNs as functions of the compactness $\xi$ are shown in Fig.~\ref{fig3}, which demonstrates that they are positive in the region $0<\xi<1$. 

For even parity, in the black-hole limit, the behavior of the TLNs for $l=2$ and $l=3$ takes the form
\begin{eqnarray}
k_2^E &\sim&\frac{40  \left(-550+1323 \tanh ^{-1}\frac{2}{5}\right)}{3 \sqrt{21} \left(-674+1625 \tanh ^{-1}\frac{2}{5}\right)}\sqrt{\xi }+\mathcal{O}\left(\xi\right)\nonumber\\
&\approx& 2.11471 \sqrt{\xi} + \mathcal{O}\left(\xi\right),
\end{eqnarray}
where the sound speed $v_s$ enters at order $\mathcal{O}\left(\xi\right)$.
\begin{eqnarray}
k_3^E &\sim& -3.2866 \xi ^{3/2} - 4.01868 v_s^2 \xi ^{3/2}+\mathcal{O}\left(\xi ^2\right).
\end{eqnarray}
The even-parity TLNs of general dS gravastars, derived from the full matching conditions, also vanish and exhibit polynomial asymptotic behavior in the BH limit, in contrast to dS gravastars with zero energy density, whose TLNs exhibit logarithmic behavior in the BH limit. 
This indicates that the asymptotic behavior of the TLNs near the BH limit is highly model dependent.

\subsection{Anti-de Sitter gravastars}
The thin-shell gravastar construction is not restricted to a de Sitter interior. Visser and Wiltshire pointed out that the same junction formalism can also be applied to configurations with an anti-de Sitter (AdS) interior, characterized by a negative cosmological constant~\cite{vdl}. Unlike conventional dS gravastars, however, AdS gravastars cannot generally be supported by a thin shell obeying the stiff equation of state assumed in the original gravastar model. Nevertheless, static configurations can be constructed by adopting other equations of state for the shell matter, including a vacuum-energy-like equation of state. Carter subsequently generalized this construction by considering both dS and AdS interiors matched to Schwarzschild--(anti-) de Sitter or Reissner--Nordstr\"om exteriors~\cite{carter}. It was shown that, for suitable shell equations of state and model parameters, AdS gravastars can be stable against spherically symmetric radial perturbations.
Closely related to these AdS thin-shell gravastars is the string-theory-inspired AdS bubble proposed by Ref.~\cite{Danielsson1}. It consists of an AdS interior and a Schwarzschild exterior separated by a thin shell composed of string-theoretic matter. Unlike more general AdS thin-shell gravastars, whose shell radius may vary as a model parameter, the equilibrium conditions for a neutral and nonrotating AdS bubble fix the shell radius at the Buchdahl value, \(r_0=9M/4\).
Previous calculations showed that the higher-multipole tidal Love numbers of this fixed-radius model become increasingly negative in the parameter region associated with instability~\cite{Ge Chen}. This sign pattern is consistent with the eikonal instability criterion for self-gravitating membranes~\cite{HY1}.

In this subsection, we consider the general thin-shell configuration with an AdS-interior/Schwarzschild-exterior geometry but treat the shell radius as a free model parameter.  We compute its odd- and even-parity tidal Love numbers using the full perturbed junction conditions and investigate their behavior as the shell approaches the black-hole limit. The background metric of AdS thin-shell gravastars is given by
\begin{equation}
\begin{cases}
	f(r) =  \left(1 - \frac{2M}{r}\right), & g(r) = f(r)^{-1},  \qquad \quad   r > r_0, \\
	f(r) = \frac{1-2M/R}{1+k R^2}(1+k r^2), & g(r) = (1+kr^2)^{-1}, \quad  r < r_0, \label{ab}
\end{cases}
\end{equation}
where $k\equiv-\frac{\Lambda}{3}>0$, and $\Lambda$ is the negative cosmological constant. Substituting the metric (\ref{ab}) into Eqs. (\ref{dp}) , the surface energy density and surface pressure of AdS gravastars are given by  
\begin{eqnarray}
	\sigma &=& \frac{1}{4 \pi r_0}(\sqrt{1+k r_0^2}-\sqrt{1-2M/r_0}),\\
	p &=& \frac{1}{8 \pi r_0}\left(\frac{1-M/r_0}{\sqrt{1-2 M/r_0}} -\frac{1+2 k r_0^2}{\sqrt{1+k r_0^2}}  \right).
\end{eqnarray}

Without loss of generality, we set \(M=1\). The TLNs depend on the parameter \(k\). To investigate their behavior in the BH limit, we consider two representative cases: \(k=10\) and \(k\to\infty\). We have also examined other values, such as \(k=1\), \(100\), and \(1000\), and found that their TLNs exhibit behavior in the BH limit similar to that for \(k=10\).

\begin{figure}[htbp]
	\centering
	\includegraphics[width=0.45\textwidth]{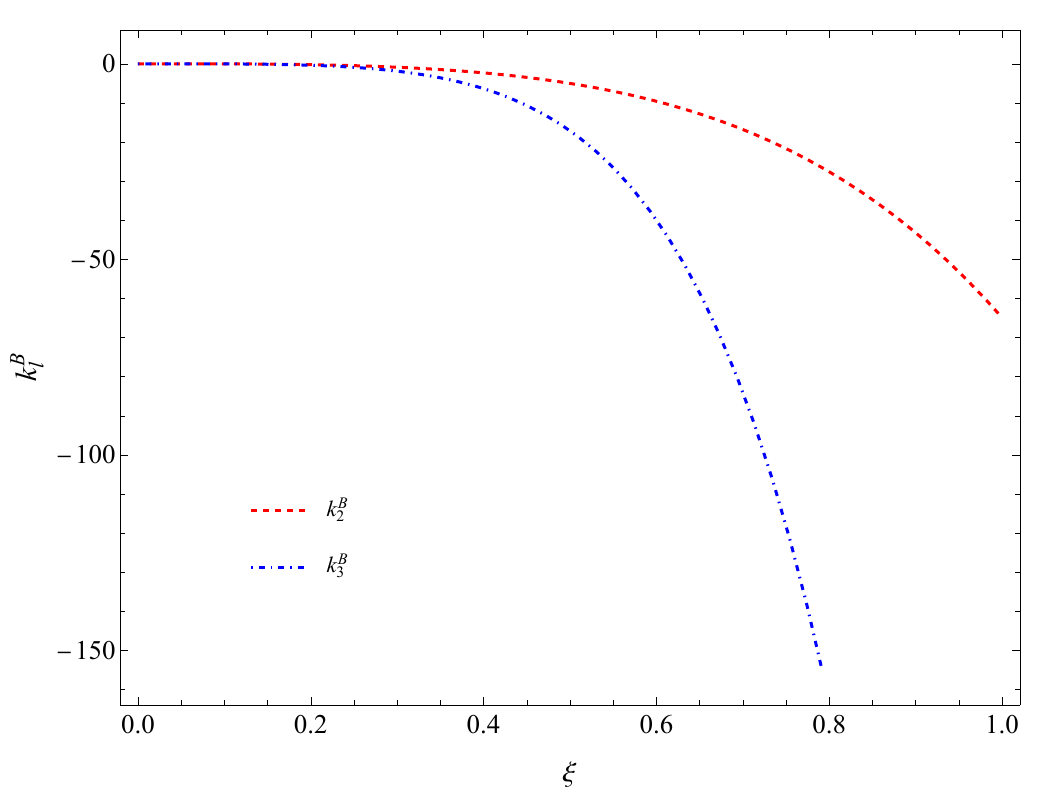}
	\caption{Odd-parity TLNs of AdS gravastars for \(l=2\) and \(l=3\) with \(k=10\). Both TLNs vanish in the BH limit and are negative throughout the range $0<\xi<1$.}
	\label{fig4}
\end{figure}

For $k=10$, the odd-parity TLNs for $l=2$ and $l=3$ are shown as functions of the compactness parameter $\xi$ in Fig.~\ref{fig4}. 
Their asymptotic behavior in the BH limit is given by
\begin{eqnarray}
	k_{2}^B\sim 1.61193 \ \xi ^{3/2} - 7.61796 \ \xi ^2+\mathcal{O}\left(\xi ^{5/2}\right),
\end{eqnarray}
\begin{eqnarray}
	k_{3}^B\sim 0.998864 \ \xi ^{3/2} - 5.88095 \ \xi ^2+\mathcal{O}\left(\xi ^{5/2}\right).
\end{eqnarray}
The odd-parity TLNs of anti-de Sitter gravastars vanish according to a power law in the BH limit, with the leading-order term scaling as $\mathcal{O}(\xi^{3/2})$. However, they remain negative throughout the range $0<\xi<1$, in contrast to those of the models discussed above. For even-parity perturbations, the TLN of \(l=2\) behaves in the BH limit as
\begin{eqnarray}
	k_{2}^E &\sim& 0.0375117 \sqrt{\xi}-0.236913\left(0.025985 + v_s^2+0.0111364 \log[\xi]\right)\xi+ \mathcal{O}(\xi ^{3/2}).
\end{eqnarray}
The even-parity TLN of anti-de Sitter gravastars also vanish as a power law in the BH limit, with the leading-order term scaling as $\mathcal{O}(\xi^{1/2})$, independently of the sound-speed parameter \(v_s^2\).

\begin{figure}[htbp]
	\centering
	\includegraphics[width=0.45\textwidth]{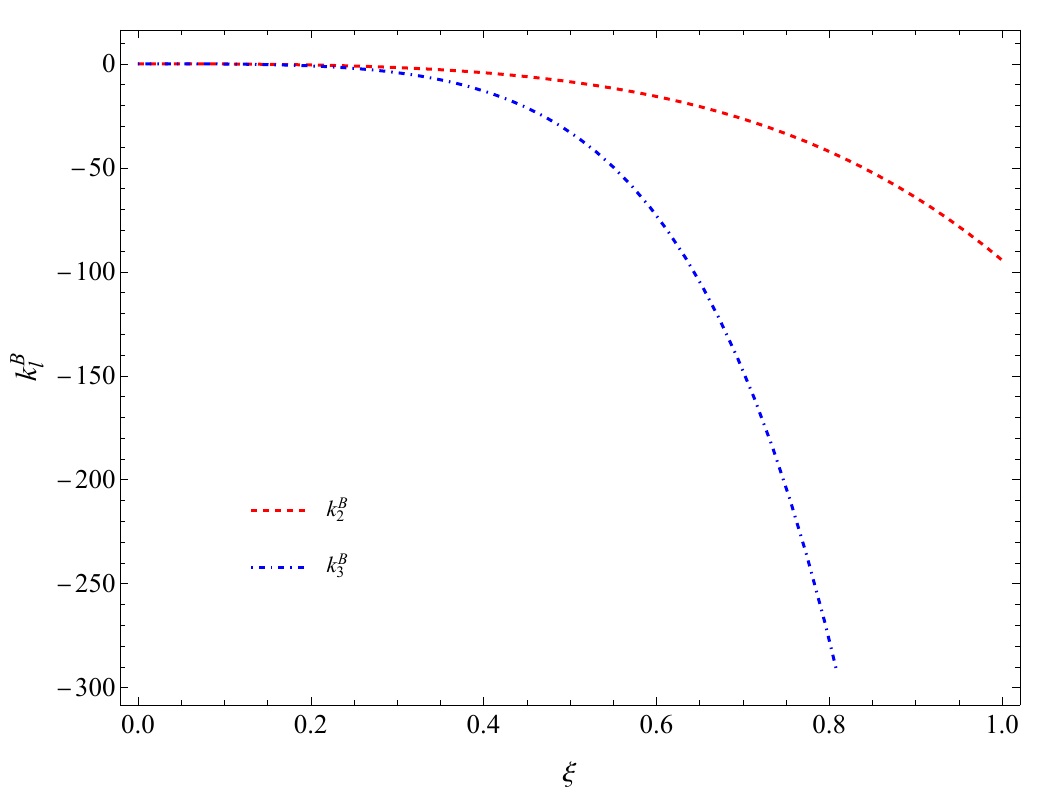}
	\caption{Odd-parity TLNs of AdS gravastars for \(l=2\) and \(l=3\) in the limit \(k\to\infty\). Both TLNs vanish in the BH limit and are negative throughout the range $0<\xi<1$.}
	\label{fig5}
\end{figure}

 We then consider the limit $k \to \infty$ and retain the leading-order contributions. Fig.~\ref{fig5} shows the odd-parity TLNs for \(l=2\) and \(l=3\) as functions of the compactness. In the BH limit, they behave as
\begin{eqnarray}
	k_{2}^B\sim -\frac{32 }{5}\xi ^2-\frac{192 }{5}\xi ^3+\mathcal{O}\left(\xi ^4\right),
\end{eqnarray}
\begin{eqnarray}
	k_{3}^B \sim -\frac{32}{7}\xi ^2-64 \xi ^3+\mathcal{O}\left(\xi ^4\right).
\end{eqnarray}
Thus, the odd-parity TLNs of anti-de Sitter gravastars vanish as a power law in the BH limit, with the leading-order term scaling as $\mathcal{O}(\xi^2)$. However, they remain negative throughout the range $0<\xi<1$, in contrast to the behavior found in the models discussed previously. For even-parity perturbations, the \(l=2\) and \(l=3\) TLNs behave in the BH limit as
\begin{eqnarray}
	k_{2}^E &\sim&	-\frac{32}{5}  \left(1+8 v_s^2\right)\xi ^2-\frac{64}{5} \left(3+48
	v_s^2 +112 v_s^4\right) \xi ^3+\mathcal{O}\left(\xi ^{7/2}\right),
\end{eqnarray}
\begin{eqnarray}
	k_{3}^E &\sim&	-\frac{32}{7}  \left(1+8 v_s^2\right)\xi ^2-\frac{64}{7}  \left(7+96
	v_s^2+208 v_s^4\right)\xi ^3+\mathcal{O}\left(\xi ^{7/2}\right).
\end{eqnarray}
Both TLNs vanish as a power law in the BH limit, with their leading-order terms scaling as \(\mathcal{O}(\xi^2)\), rather than exhibiting logarithmic behavior. The signs of the even-parity TLNs depend on the sound-speed parameter $v_s^2$; in particular, both TLNs are negative for normal matter with $0<v_s^2<1$.

Our results also show that, for AdS gravastars, the leading-order contributions become progressively suppressed as \(k\) increases. In the limit \(k\to\infty\), the leading nonvanishing term scales as \(\xi^2\).

\subsection{Comparison of the perturbed junction conditions}
To clarify the origin of the discrepancies between our results and those reported in Ref.~\cite{Vitor}, we first review the two approaches used to formulate the perturbed Israel junction conditions and establish the correspondence between them.

A physically consistent matching of the interior and exterior perturbed spacetimes at the location of the perturbed shell requires the imposition of Israel’s junction conditions. This issue is delicate and has been addressed mainly through two approaches~\cite{Cardoso1, Paolo1}. 
One of these approaches was proposed in 2009 in Ref.~\cite{Paolo1}. Ref.~\cite{Paolo1} argued that, technically, the application of Israel’s junction conditions is easier if the shell’s world tube coincides with a coordinate sphere at a fixed radius. However, this requirement is incompatible with choosing the Regge-Wheeler (RW) gauge in both the interior and exterior spacetimes, although this gauge is convenient for casting the perturbation equations into simpler forms. In other words, the Regge-Wheeler gauge is incompatible with the requirement that the shell’s world tube remain at a fixed coordinate location. In order to take advantage of both the simplicity of the field equations and the convenience of matching on a fixed coordinate sphere, they carry out the matching in the following way. They first construct a particular coordinate transformation (for both the exterior and interior spacetimes) such that in the new coordinate system, any mass on the shell remains static on the coordinate sphere with $r = r_0$. In this new coordinate system, the metric perturbations will no longer be in the RW gauge, but will be augmented by quantities that carry information about how masses on the shell move in the RW gauge. The stress-energy tensor of masses on the shell will correspondingly be modified. They then carry out the matching at $r = r_0$ and obtain junction conditions relating the interior and exterior metric perturbations, together with equations of motion for matter on the shell. The second approach does not relocate the perturbed shell to a fixed coordinate sphere. Instead, Israel's junction conditions are imposed directly at the actual perturbed location of the shell~\cite{Cardoso1}. These two approaches are essentially different descriptions of the same geometric matching problem. In the first approach, a coordinate transformation is performed to fix the shell at \(r=r_0\), thereby transferring information about the shell's displacement to the transformed metric perturbations and surface stress-energy tensor. In the second approach, the shell displacement is retained explicitly, and the matching is performed directly on the perturbed hypersurface. Therefore, provided that all first-order contributions from the shell displacement, metric perturbations, and perturbations of the surface matter are fully retained and that the gauge transformations are treated consistently, the physical junction conditions obtained using the two approaches should ultimately be equivalent. For convenience, we establish a correspondence between the perturbed junction conditions obtained using the two approaches. Note that Ref.~\cite{Paolo1} considers general dynamical perturbations, whereas we focus on the static case. Therefore, the junction conditions presented in our manuscript do not contain time-derivative terms. Moreover, some of the even-parity junction conditions
have been substantially simplified in Ref.~\cite{Paolo1}, whereas our manuscript presents them in their original, unsimplified form. Consequently, some of the corresponding expressions may appear different. For odd-parity perturbations, the first junction condition yields Eq.~(B3) of our manuscript, while the second junction condition yields the first equation of Eq.~(B6) and Eq.~(B7). In Ref.~\cite{Paolo1}, the first junction condition yields the fourth equation of Eq.~(A20), which is equivalent to Eq.~(B3). The second junction condition yields Eq.~(A35), which corresponds to the first equation of Eq.~(B6). However, Ref.~\cite{Paolo1} does not explicitly present the result corresponding to Eq.~(B7), which can be derived by combining Eqs.~(A32) and (A26) of Ref.~\cite{Paolo1}. 
For even-parity perturbations, the first junction condition yields Eq.~(B4) of our manuscript, while the second junction condition yields Eq.~(B5), the second equation of Eq.~(B6), and Eqs.~(B8) and (B9). In Ref.~\cite{Paolo1}, the first junction condition yields the fifth equation in Eq.~(A20), or equivalently Eq.~(A39), which corresponds to the first equation of Eq.~(B4). It also yields the sixth equation of Eq.~(A20), or equivalently Eq.~(A40), which corresponds to the second equation of Eq.~(B4). The second junction condition yields Eq.~(A36), corresponding to Eq.~(B5); Eq.~(A37), corresponding to Eq.~(B8); and Eq.~(A38), corresponding to Eq.~(B9). Ref.~\cite{Paolo1} does not explicitly provide the expression involving $H_1$, which corresponds to the second equation of Eq.~(B6), but it can be derived by combining Eqs.~(A25) and (A31) of Ref.~\cite{Paolo1}. The junction conditions given in Eqs.~(A37) and (A38) of Ref.~\cite{Paolo1} are substantially simplified, whereas Eqs.~(B8) and (B9) presented in our manuscript retain their original form. Although these two expressions may appear different, they are equivalent forms of the same junction conditions. 

Based on the discussion above, let us comment on the junction conditions used in Ref.~\cite{Vitor}. Although Ref.~\cite{Vitor} claims to adopt the formalism developed in Ref.~\cite{Paolo1}, it does not apply the correct perturbed junction conditions.
For odd-parity perturbations, there are two independent junction conditions. The first is $[[h_0]]=0$, and the second concerns $h_0^{'}$. Ref.~\cite{Paolo1} did not explicitly list the condition on $h_0^{'}$, which arises from the second junction condition and can be derived from Eqs.~(A32) and (A26) of Ref.~\cite{Paolo1}. Ref.~[48] imposed the condition $h_0^{'}=0$ without justification, which violates the second junction condition. For even-parity perturbations, in the Mazur–Mottola gravastar with $\Sigma=0$,
the junction condition (A36) of Ref.~\cite{Paolo1} implies $[[z]]=0$, which is equivalent to $ [[\delta r]] = 0$. Therefore, $[[\delta r]]=0 $ is valid only for the special case of the Mazur–Mottola gravastar with $\Sigma=0$ and does not generally hold for other thin-shell models. The conditions derived from the second junction condition, Eqs.~(A36)–(A40) of Ref.~\cite{Paolo1}, reduce to Eqs.~(3.30)–(3.32) of the same reference. These equations provide three equations for four unknown constants: three integration constants associated with the metric perturbation $H$ and one constant associated with the surface-density perturbation $\delta\Sigma$. The tidal Love number can therefore be determined, and the resulting expression depends on the squared sound speed $v_s^2$. However, for simplicity, Ref.~\cite{Vitor} further assumed that the thin-shell material is stiff, so that $\delta\Sigma \sim 0$. Equation (3.31) then reduces to \([[K']]=0\). Therefore, the junction conditions become \([[K]]=0\) and \([[K']]=0\), which are given in Eq.~(26) of our manuscript. The condition \([[\delta r]]=0\) was not imposed as an additional assumption; rather, it is automatically satisfied in the special case of the Mazur–Mottola gravastar with \(\Sigma=0\). 
For wormholes, however, \(\Sigma\neq 0\), so \([[K]]=0\) cannot be derived from Eq.~(A39). Therefore, imposing \([[K]]=0\) violates the first junction condition.
Eq.~(3.31) is derived from Eqs.~(A37) and (A40) under the assumption that \(\Sigma=0\), which is not satisfied by the wormhole model. Ref.~\cite{Vitor} further assumes that the thin-shell matter is stiff, so that \(\delta\Sigma \sim 0\), and then uses Eq.~(3.31) to obtain \([[K']]=0\). However, because Eq.~(3.31) is not applicable to the wormhole model, this reasoning is invalid. Therefore, the condition \([[K']]=0\) imposed for wormholes in Ref.~\cite{Vitor} also violates the second junction condition. 

In summary, we have carefully examined the perturbed junction conditions proposed in Refs.~\cite{Cardoso1, Paolo1}. We find that the odd-parity junction conditions used in Ref.~\cite{Vitor} are incorrect for both wormholes and de Sitter gravastars with zero surface energy density. The even-parity junction conditions used for wormholes are also incorrect, while the even-parity junction conditions used for de Sitter gravastars are valid only in the special case of zero surface energy density and under the assumption of stiff shell matter, \(\delta\Sigma\sim 0\).

\section{Conclusions and Summary}

Tidal Love numbers provide an important probe of the internal structure and gravitational response of compact objects in external tidal environments. With the rapid development of gravitational-wave astronomy, TLNs have become a powerful tool for distinguishing black holes from exotic compact objects and for testing gravity in the strong-field regime. Since black holes in classical general relativity possess vanishing TLNs, any detectable nonzero tidal deformability could provide evidence for new physics beyond the standard black-hole paradigm. Understanding the behavior of TLNs for different ECO models is therefore of considerable theoretical and observational importance.

In this work, we revisited the tidal Love numbers of several classes of exotic compact objects by employing the full perturbed junction conditions recently derived in Ref.~\cite{Cardoso1}. In particular, we considered four representative thin-shell models: wormholes, de Sitter gravastars with zero surface energy density, general de Sitter gravastars, and anti-de Sitter gravastars. Our main goal was to investigate how the full perturbed matching conditions modify the behavior of the TLNs relative to the results obtained previously using simplified junction conditions~\cite{Vitor}. Previous studies showed that the TLNs of ECOs are generically nonvanishing, although they approach zero in the black-hole (BH) limit. For thin-shell wormholes and gravastars with zero energy density, Ref.~\cite{Vitor} found that both even- and odd-parity TLNs vanish logarithmically in the limit $r_0 \rightarrow 2M$, leading to the conjecture that logarithmic behavior is a universal feature of ultracompact ECOs whose exterior geometry approaches that of a Schwarzschild BH. Our results show that this conjecture is not generally valid once the full perturbed junction conditions are taken into account. In particular, we find that the odd-parity TLNs do not generically exhibit logarithmic behavior in the BH limit. Instead, their asymptotic behavior becomes polynomial in the compactness parameter $ \xi $. More specifically, for thin-shell wormholes and gravastars with zero surface energy density, the odd-parity TLNs scale as $ \mathcal{O}(\xi) $. For general dS gravastars, the leading behavior becomes $\mathcal{O}(\xi^{3/2})$, while for anti-de Sitter gravastars the odd-parity TLNs scale as $\mathcal{O}(\xi^2)$. Therefore, the logarithmic behavior previously conjectured to be universal is violated in a broad class of models when the complete perturbed junction conditions are imposed. For even-parity perturbations, the situation is more involved because the perturbations couple directly to the shell matter degrees of freedom and depend sensitively on the square of the sound speed $v_s^2$. We find that the TLNs of thin-shell wormholes and dS gravastars with zero energy density still exhibit logarithmic behavior in the BH limit, indicating that the logarithmic scaling survives in certain special cases. However, this behavior is again not universal. For general dS gravastars, the even-parity TLNs scale as $ \mathcal{O}(\xi^{1/2}) $, whereas for anti-de Sitter gravastars the leading behavior becomes $ \mathcal{O}(\xi^2) $. We also analyzed the signs of the TLNs in detail. In the odd-parity sector, the TLNs of thin-shell wormholes and dS gravastars with zero energy density are positive, in contrast with the negative values reported previously in Ref.~\cite{Vitor}. The odd-parity TLNs of general dS gravastars are likewise positive, while those of anti-de Sitter gravastars remain negative. In the even-parity sector, the signs of the TLNs depend strongly on the shell matter properties. For wormholes and dS gravastars with zero energy density, the logarithmic behavior implies negative TLNs. For general dS gravastars, we find that $k_2^E$ is positive, whereas the sign of $k_3^E$ depends sensitively on the sound speed $v_s^2$; for ordinary matter satisfying $0< v_s^2<1$, $k_3^E$ is negative. Similarly, for anti-de Sitter gravastars both $k_2^E$ and $k_3^E$ are closely related to $v_s^2$ and are negative for normal matter.

Overall, our analysis demonstrates that the full perturbed junction conditions can qualitatively modify both the asymptotic behavior and the signs of the TLNs of exotic compact objects. In particular, the previously conjectured universal logarithmic scaling of TLNs near the BH limit is not generic. These results suggest that the tidal response of ultracompact objects is considerably more sensitive to the detailed matching conditions and shell dynamics than previously expected. In particular, Refs.~\cite{HY1,tp,Ge Chen} show that the negativity of even-parity tidal Love numbers is directly associated with the existence of unstable normal modes. Therefore, our results may have important implications for the dynamical instability of ECOs.

\begin{acknowledgments}
This work is partially supported by the National Key Research and Development
Program of China with Grant No.~2021YFC2203001 as well as the National Natural Science Foundation of China with Grant Nos.~12361141825, 12375058, 12035016, 12475049, 11647090, and 12575047.

\end{acknowledgments}

\newpage
\appendix	
\section{Solutions for metric perturbations}\label{solution}
In this appendix, we present the explicit expressions for the metric perturbations. The detailed derivation of these solutions can be found in the literature~\cite{Vitor,Emanuele Berti}.
\subsection{Exterior solution}
The exterior spacetime of our models is described by the Schwarzschild metric. For even parity, a static tidal field implies $H_1=0$ and $H_2=H_0\equiv H$.
\begin{eqnarray}
	H^{\text{out}}(r)=c_1 \hat{P}_{\ell 2}(x) + c_2 \hat{Q}_{\ell 2}(x)|_{x\equiv r/M-1},\label{evenout}
\end{eqnarray}
where $c_1$ and $c_2$ are integration constants determined by matching to the interior solution, and $\hat{P}_{\ell 2}(x)$ and $\hat{Q}_{\ell 2}(x)$ are normalized associated Legendre functions, explicitly given by
\begin{eqnarray}
	\hat{P}_{\ell 2}(x)&=&\frac{\sqrt{\pi } 2^{-\ell-1} \left(x^2-1\right) \Gamma (\ell+3) \, _2F_1\left(2-\ell,\ell+3;3;\frac{1-x}{2}\right)}{\left(4 \ell^2-8 \ell+3\right)
		\Gamma \left(\ell-\frac{3}{2}\right)},\\
	\hat{Q}_{\ell 2}(x)&=&\left(x^2-1\right) x^{-\ell-3} \, _2F_1\left(\frac{\ell+3}{2},\frac{\ell+4}{2};\ell+\frac{3}{2};\frac{1}{x^2}\right),
\end{eqnarray}
where $_2F_1$ denotes the hypergeometric function. The metric perturbation function $K(r)$ is given by
\begin{eqnarray}
	K(r)=\frac{2 M H'(r)}{\ell ^2+\ell -2}+\frac{H(r) \left [4 M^2+2 M r \left(\ell ^2+\ell -4\right)-r^2 \left(\ell ^2+\ell
		-2\right)\right]}{r \left(\ell ^2+\ell -2\right) (2 M-r)}.
\end{eqnarray}

For odd parity, $h_1=0$, and
\begin{eqnarray}
	h_{0}(r)&=&\tilde{c}_{1}(\frac{r}{ M})^{\ell +1}\,   _2F_1\left(-\ell +1,-\ell-2;-2\ell;\frac{2 M}{r}\right)\nonumber\\
	&+&\tilde{c}_{2}(\frac{ M}{r})^\ell \,   _2F_1\left(\ell-1,\ell+2;2\ell+2;\frac{2M}{r}\right),\label{oddout}
\end{eqnarray}
where $\tilde{c}_1$ and $\tilde{c}_2$ are integration constants determined by matching to the interior solution.

\subsection{Interior solution}

The interior regions of our models are different, and therefore the corresponding perturbation solutions must be discussed separately.

\subsubsection{Wormholes}
For wormholes, the interior spacetime is an exact copy of the exterior metric. Since there are no tidal fields in the interior region, the even-parity solution is given by
\begin{eqnarray}
	H^{\text{in}}(r)= c_2 \hat{Q}_{\ell 2}(x)|_{x\equiv r/M-1}.
\end{eqnarray}
The metric perturbation function $K(r)$ is given by
\begin{eqnarray}
	K(r)=\frac{2 M H'(r)}{\ell ^2+\ell -2}+\frac{H(r) \left [4 M^2+2 M r \left(\ell ^2+\ell -4\right)-r^2 \left(\ell ^2+\ell
		-2\right)\right]}{r \left(\ell ^2+\ell -2\right) (2 M-r)}.
\end{eqnarray}

For odd parity, the absence of tidal fields implies
\begin{eqnarray}
	h_{0}(r)=\tilde{c}_{3}(\frac{M}{r})^\ell \,   _2F_1\left(\ell-1,\ell+2;2\ell+2;\frac{2M}{r}\right).
\end{eqnarray}

\subsubsection{De Sitter gravastars}
For even parity,  $H_1=0$ and $H_2=H_0\equiv H$, with
\begin{eqnarray}
	H(r)=c_{3}  \frac{\sqrt{\Lambda} r ^{\ell } }{3-\Lambda  r ^2}\, _2F_1\left(\frac{\ell -1}{2},\frac{\ell }{2};\ell +\frac{3}{2};\frac{\Lambda  r
		^2}{3}\right),\label{ccd1}
\end{eqnarray}
where $c_3$ is an integration constant. The metric perturbation function $K(r)$ is given by
\begin{eqnarray}
	K(r)=\frac{H(r ) \left[-2 \left(\Lambda ^2 r ^4+6 \Lambda  r^2-9\right)+3 \ell ^2
		\left(\Lambda  r^2-3\right)+3 \ell  \left(\Lambda  r^2-3\right)\right]-2 \Lambda
		r^3 \left(\Lambda  r^2-3\right) H'(r)}{3 \left(\ell ^2+\ell -2\right)
		\left(\Lambda  r^2-3\right)}.
\end{eqnarray}

For odd parity, $h_1=0$, and 
\begin{eqnarray}
	h_{0}(r)=\tilde{c}_{3}\Lambda^{\frac{\ell +1}{2}}r^{\ell +1} \, _2F_1\left(\frac{\ell
	}{2}-\frac{1}{2},\frac{\ell }{2}+1;\ell
	+\frac{3}{2};\frac{\Lambda  }{3}r^2\right),\label{2337}
\end{eqnarray}
where $\tilde{c}_3$ is an integration constant.

\subsubsection{Anti-de Sitter gravastars}
For even parity,  $H_1=0$ and $H_2=H_0\equiv H$, with
\begin{eqnarray}
	H(r)=c_{3} \sqrt{-\Lambda} \frac{r ^{\ell } \, _2F_1\left(\frac{\ell -1}{2},\frac{\ell }{2};\ell +\frac{3}{2};\frac{\Lambda  r
			^2}{3}\right)}{3-\Lambda  r ^2},\label{ccd1}
\end{eqnarray}
where $c_3$ is an integration constant. The metric perturbation function $K(r)$ is given by
\begin{eqnarray}
	K(r)=\frac{H(r ) \left[-2 \left(\Lambda ^2 r ^4+6 \Lambda  r^2-9\right)+3 \ell ^2
		\left(\Lambda  r^2-3\right)+3 \ell  \left(\Lambda  r^2-3\right)\right]-2 \Lambda
		r^3 \left(\Lambda  r^2-3\right) H'(r)}{3 \left(\ell ^2+\ell -2\right)
		\left(\Lambda  r^2-3\right)}.
\end{eqnarray}

For odd parity, $h_1=0$, and 
\begin{eqnarray}
	h_{0}(r)=\tilde{c}_{3}(-\Lambda)^{\frac{\ell +1}{2}}r^{\ell +1} \, _2F_1\left(\frac{\ell
	}{2}-\frac{1}{2},\frac{\ell }{2}+1;\ell
	+\frac{3}{2};\frac{\Lambda  }{3}r^2\right),\label{2337}
\end{eqnarray}
where $\tilde{c}_3$ is an integration constant.

\section{The junction conditions for perturbed shells}\label{jc}

To match the interior and exterior  perturbed spacetimes physically at the location of the shell, we need to impose Israel's junction conditions~\cite{Israel,Israel2,Eric}. This issue is subtle and has been addressed in several studies~\cite{Paolo1,Uchikata1,Cardoso1}. Recently, Ref.~\cite{Cardoso1} derived the full perturbed junction conditions. Here, we summarize  these results.

The shell radius is perturbed by an external tidal field
\begin{eqnarray}
	\delta r_{\pm}=\delta r^{l0}_{\pm}Y^{l0}(\Theta,\Phi),
\end{eqnarray}
where we set $m=0$ in the spherical-harmonic decomposition. The perturbed shell is therefore located at
\begin{eqnarray}
	\left( T,r_0+\delta r_{\pm}Y^{l0}(\Theta,\Phi),\Theta,\Phi \right),
\end{eqnarray}
where we omit the harmonic index $l0$ in $\delta r^{l0}_{\pm}$ hereafter to avoid clutter.

We consider only linear perturbations. The first junction condition gives 
\begin{eqnarray}
	[[h_0]]=0, \label{1jc}
\end{eqnarray}
\begin{eqnarray}
 \frac{2}{r_0}[[\delta r]]=-[[K]], \; \; [[H_0]]=[[\frac{\delta r f^{\prime}}{f}]]. \label{fjc2}
\end{eqnarray}

The second junction condition gives \footnote{In equation (72) of the reference~\cite{Cardoso1},
	the coefficients of the first and third terms should both be $1$. We have corrected these typos in equation (\ref{even6}).}
\begin{eqnarray}
	[[\sqrt{g}\delta r]]=0,\label{even1}
\end{eqnarray}
\begin{eqnarray}
 [[\frac{h_1}{\sqrt{g}} ]]=0,  \;  \;  [[\frac{H_1}{\sqrt{g}}]]=0.
\end{eqnarray}

\begin{eqnarray}
	\frac{1}{2}[[\frac{h^\prime_{0}}{\sqrt{g}}]]-\frac{2}{r_0}[[\frac{1}{\sqrt{g}}]]h_0
	-\frac{1}{2}[[\frac{f^{\prime}}{f\sqrt{g}}]]h_0=8\pi \sigma h_0.\label{odd2}
\end{eqnarray}
\begin{eqnarray}
	\frac{2}{r_0^2}[[\frac{\delta r}{\sqrt{g}}]]+\frac{2}{r_0}[[\frac{H_0}{\sqrt{g}}]]+
	\frac{1}{r_0}[[\frac{H_2}{\sqrt{g}}]]-[[\frac{K^{\prime}}{\sqrt{g}}]]+\frac{1}{r_0}[[\frac{\delta r g^{\prime}}{\sqrt{g^3}}]]\nonumber\\
	-\frac{2}{r_0}[[\frac{\delta r f^{\prime}}{f\sqrt{g}}]]=8\pi \delta \sigma + 8\pi \sigma(\frac{f^{\prime}\delta r}{f} -H_0).\label{even2}
\end{eqnarray}
\begin{eqnarray}\label{even6}
	&&\frac{1}{r_0^2}[[\frac{\delta r}{\sqrt{g}}]]-\frac{1}{2r_0}[[\frac{H_2}{\sqrt{g}}]]+\frac{1}{r_0}[[\frac{K}{\sqrt{g}}]]-\frac{1}{4}[[\frac{H_2f^{\prime}}{f\sqrt{g}}]]
	+\frac{1}{2}[[\frac{Kf^{\prime}}{f\sqrt{g}}]]\nonumber\\
	&&+\frac{1}{2}[[\frac{K^{\prime}}{\sqrt{g}}]]-\frac{1}{2}[[\frac{H_0^{\prime}}{\sqrt{g}}]]-\frac{1}{2 r_0}[[\frac{\delta r g^{\prime}}{\sqrt{g^3}}]]+\frac{1}{r_0}[[\frac{\delta r f^{\prime}}{f\sqrt{g}}]]+
	\frac{1}{2}[[\frac{\delta r f''}{f\sqrt{g}}]]\nonumber\\
	&&-\frac{1}{2}[[\delta r (\frac{f'}{f})^2\frac{1}{\sqrt{g}}]]
	-\frac{1}{4}[[\delta r \frac{f'}{f}\frac{g'}{g^{3/2}}]]=8\pi \delta p+8\pi p(K+2\frac{\delta r}{r_0}). \label{even3}
\end{eqnarray}

In the odd-parity sector, $h_1$ vanishes for static perturbations, leaving two junction conditions for three constants of integration coming from $h_0$ (one from inside the shell and two
from outside). Since an overall factor is irrelevant to the computation of the TLNs, we already possess sufficient information to fully determine the odd-parity TLNs. For even parity, a complete description requires the addition of an equation of state relating $\delta p$ to $\delta\sigma$
\begin{eqnarray}
	\delta p=\upsilon^2_s\delta\sigma,\; \;  \upsilon^2_s\equiv(\frac{dp}{d\sigma})|_{\sigma_0},
\end{eqnarray}
where $\upsilon_s$ is the sound speed of the fluid. 
For static perturbations, $H_1$ vanishes, leaving five junction conditions for six constants of integration (three from $H$, two from $\delta r$, and one from
$\delta \sigma$). As a result, five constants can be determined, leaving a single free constant which does not affect the computation of the even-parity TLNs.


\begin{thebibliography}{99}
	\bibitem{Wald} R. M. Wald, General Relativity (University of Chicago press, USA, 1984).
	
	\bibitem{Chandrasekhar} S. Chandrasekhar, The mathematical theory of black holes (Oxford university press, 1998).
	
	
	\bibitem{B1}B. P. Abbott et al. (LIGO Scientific Collaboration and Virgo Collaboration), Phys. Rev. Lett. \textbf{116}, 061102 (2016) .
	
	
	\bibitem{B2}B. P. Abbott et al. and Abbott. (LIGO Scientific Collaboration and Virgo Collaboration), Phys. Rev. X \textbf{9}, 031040 (2019).
	\bibitem{R1}R. Abbott et al.  (LIGO Scientific Collaboration and Virgo Collaboration), Phys. Rev. X \textbf{11}, 021053 (2021) .
	\bibitem{R2} R. Abbott et al. (LIGO Scientific Collaboration, Virgo Collaboration, and KAGRA Collaboration), Phys. Rev. X \textbf{13},
	041039 (2023).
\bibitem{AG}  A. G. Abac et al. (LIGO Scientific Collaboration, Virgo Collaboration, and KAGRA Collaboration), arXiv:2508.18082 [gr-qc].
	
	\bibitem{S}  S. Doeleman et al., arXiv:0906.3899 [astro-ph.CO].
	
	\bibitem{K}  K. Akiyama et al., Astrophys. 
	J. Lett. \textbf{875}, L1 (2019) .
	
\bibitem{RS}R. Sch\"odel et al., Nature \textbf{419}, 694–696 (2002) .
	
	\bibitem{Hawking1}S. W. Hawking, Phys. Rev. D \textbf{14}, 2460 (1976) .
	
	
	\bibitem{V. Cardoso}V. Cardoso and P. Pani, Living Reviews in Relativity 22
	(2019).
	
	\bibitem{M. Visser}M. Visser, C. Barcelo, S. Liberati, and S. Sonego, arXiv:0902.0346
	[gr-qc] .
	\bibitem{N. K} N. K. Johnson-McDaniel, A. Mukherjee, R. Kashyap, P. Ajith, W. Del Pozzo, and S. Vitale, Phys. Rev. D \textbf{102}, 123010 (2020).
		
\bibitem{Berti} E.  Berti,  A.  Sesana,  E.  Barausse,  V.  Cardoso,  and K. Belczynski, Phys. Rev. Lett. \textbf{117}, 101102 (2016).

\bibitem{Yang}H. Yang, K. Yagi, J. Blackman, L. Lehner, V. Paschalidis, F. 
Pretorius, and N. Yunes, Phys. Rev. Lett. \textbf{118}, 161101(2017).

	
	
	
	
	
	
	\bibitem{Jetzer1}P. Jetzer, Phys. Rept. \textbf{220}, 163 (1992).
	
	\bibitem{Jetzer2}F. E. Schunck and E. W. Mielke, Classical and Quantum Gravity
	\textbf{20}, 301 (2003).
	
	\bibitem{Jetzer3}S. L. Liebling and C. Palenzuela, Living Rev. Relativity \textbf{26}, 1
	(2023).
	
	
	\bibitem{Finster}F. Finster, J. Smoller, and S. T. Yau, Phys. Rev. D \textbf{59},
	104020 (1999).
	
	
	\bibitem{Brito}R. Brito, V. Cardoso, C. A. R. Herdeiro, and E. Radu,
	Phys. Lett. B \textbf{752}, 291 (2016).
	
	\bibitem{Lee}T. D. Lee and Y. Pang, Phys. Rev. D \textbf{35}, 3678 (1987).
	
	\bibitem{Grosso}L. D. Grosso, G. Franciolini, P. Pani, and A. Urbano, Phys. Rev. D \textbf{108}, 044024 (2023).
	
	
	\bibitem{Grosso1}L. D. Grosso and P. Pani, Phys. Rev. D \textbf{108}, 064042
	(2023).
	
	\bibitem{ms}M.  Visser,  Lorentzian wormholes:  From Einstein to 
	Hawking (AIP, Woodbury, New York, 1996) .
	
	\bibitem{POM}P. O. Mazur and E. Mottola, Universe, \textbf{9}, 88 (2023).
	
	\bibitem{mdl} M. Visser and D. L. Wiltshire, Classical Quantum Gravity
	\textbf{21}, 1135 (2004) .
	
	
	\bibitem{Kalogera}V. Kalogera et al., arXiv:2111.06990 [gr-qc] (2021).
	\bibitem{Maggiore}M. Maggiore et al., Journal of Cosmology and Astroparticle Physics \textbf{03}, 050 (2020).
	
	\bibitem{Branchesi} M. Branchesi et al., Journal of Cosmology and Astroparticle Physics,  \textbf{07}, 068 (2023).
	
	\bibitem{Abbott}B. P. Abbott et al. (LIGO Scientific), Classical and Quantum Gravity
	\textbf{34}, 044001 (2017).
	
	\bibitem{Essick}R. Essick, S. Vitale, and M. Evans, Phys. Rev. D \textbf{96},
	084004 (2017).
	
	\bibitem{Pacilio}C. Pacilio, A. Maselli, M. Fasano, and P. Pani, Phys.
	Rev. Lett. \textbf{128}, 101101 (2022).
	
	\bibitem{Forteza}X. J. Forteza, T. Abdelsalhin, P. Pani, and
	L. Gualtieri, Phys. Rev. D \textbf{98}, 124014 (2018).
	
	\bibitem{Love}A. E. H. Love,
	 Mont. Not. Roy. Astr. Soc. \textbf{69}, 476 (1909).
	
	\bibitem{Murray} C. Murray and S. Dermott, Solar System Dynamics (Cambridge University Press, Cambridge, UK, 2000).
	
	\bibitem{Will}E. Poisson and C. Will, Gravity: Newtonian, post-newtonian, relativistic (Cambridge University Press, Cambridge, UK, 2014).
	
	\bibitem{Hinderer} T. Hinderer, Astrophys. J. \textbf{677}, 1216 (2008). [Erratum: Astrophys. J. \textbf{697}, 964 (2009)].
	
	\bibitem{Poisson} T. Binnington and E. Poisson, Phys. Rev. D \textbf{80}, 084018 (2009).
	\bibitem{Damour} T. Damour and A. Nagar, Phys. Rev. D \textbf{80}, 084035 (2009).
	
	\bibitem{Lattimer1} J. Lattimer and M. Prakash, Science \textbf{304}, 536 (2004).
	
	\bibitem{Lattimer2} T. Hinderer, B. D. Lackey, R. N.  Lang, and J. S. Read,
	Phys. Rev. D \textbf{81}, 123016 (2010).
	
	\bibitem{Lattimer3}S. Postnikov, M. Prakash, and J. M. Lattimer, Phys. Rev. D \textbf{82}, 024016 (2010).
	
	\bibitem{Lattimer4} J. Vines, E. E. Flanagan, and T. Hinderer, Phys. Rev. D \textbf{83}, 084051 (2011).
	
	\bibitem{Lattimer5} T. Damour, A. Nagar, and L. Villain, Phys. Rev. D \textbf{85},
	123007 (2012).
	
	\bibitem{Lattimer6} W. D. Pozzo, T. G. F. Li, M. Agathos, C. V. D. Broeck, and S. Vitale, Phys. Rev. Lett. \textbf{111}, 071101
	(2013).
	
	\bibitem{Lattimer7}A. Maselli, L. Gualtieri, and V. Ferrari, Phys. Rev. D \textbf{88}, 104040 (2013).
	
	\bibitem{Vitor}V. Cardoso, E. Franzin, A. Maselli, P. Pani, G. Raposo, Phys. Rev. D \textbf{95}, 084014 (2017).
	
	\bibitem{Cardoso1}V. Cardoso and F. Duque, Phys. Rev. D \textbf{101}, 064028 (2020).
	
		\bibitem{Israel} W. Israel, Il Nuovo Cimento B (1965-1970) \textbf{44}, 1 (1966).
	
	\bibitem{Israel2} C. Barrabes and W. Israel, Phys. Rev. D \textbf{43}, 1129 (1991).
	
	\bibitem{Eric} E. Poisson, A relativist's toolkit: the mathematics of black-hole mechanics (Cambridge University Press, 2004).
	

	
	\bibitem{Wheeler}T. Regge and J. A. Wheeler, Physical Review \textbf{108}, 1063 (1957).
	
	\bibitem{Geroch}R. P. Geroch, J. Math. Phys. \textbf{11}, 2580 (1970).
	
	\bibitem{Hansen}R. O. Hansen, J. Math. Phys. \textbf{15}, 46 (1974).
	
	\bibitem{Thorne} K. S. Thorne, Reviews of Modern Physics \textbf{52}, 299 (1980).
	
	\bibitem{YG}Y. G{\"u}rsel, General relativity and gravitation \textbf{15}, 737 (1983).
	
	
	\bibitem{Visser} M. Visser, Lorentzian wormholes: From Einstein to
	Hawking (AIP, Woodbury, New York, 1996).
	

	
	\bibitem{Paolo1}P. Pani, E. Berti, V. Cardoso, Y. Chen, and R. Norte, Phys. Rev. D \textbf{80}, 124047 (2009).
	
	

	
	

	
	
\bibitem{po}P. O. Mazur and E. Mottola, Universe \textbf{9}, 88 (2023) .


\bibitem{vdl} M. Visser and D. L. Wiltshire, Classical Quantum Gravity
\textbf{21}, 1135 (2004) .



\bibitem{nk} N. K. Johnson-McDaniel, A. Mukherjee, R. Kashyap, P. Ajith, W. Del Pozzo, and S. Vitale, Phys. Rev. D \textbf{102}, 123010 (2020) .

\bibitem{ve} V. Cardoso, E. Franzin, and P. Pani, Phys. Rev. Lett. \textbf{116}, 171101 (2016) .


\bibitem{Uchikata1}N. Uchikata and S. Yoshida, Classical and Quantum Gravity \textbf{33}, 025005 (2015).

\bibitem{carter}B.~M.~N.~Carter,
Class. Quantum Grav. \textbf{22}, 4551 (2005).

\bibitem{Danielsson1}U. H. Danielsson, G. Dibitetto and S. Giri, Journal of High Energy Physics \textbf{2017}, 171, (2017)	

\bibitem{Ge Chen}G. Chen, Y. Tian, and H. Zhang, Phys. Rev. D \textbf{113}, 124057 (2026).

\bibitem{HY1}H. Yang, B. Bonga, and Z. Pan, Phys. Rev. Lett. \textbf{130}, 011402 (2023).



\bibitem{tp}T. Pitre, B. Schneider, and E. Poisson, Phys. Rev. D \textbf{113}, 124055 (2026).

	
\bibitem{Emanuele Berti}E. Berti, V. D. Luca, L. D. Grosso, and P. Pani, Phys. Rev. D \textbf{109}, 124008 (2024).




	
\end{thebibliography}
\end{document}